\documentclass{aa}

\usepackage[colorlinks, citecolor=blue]{hyperref}
\usepackage{pdflscape}
\usepackage{subcaption}
\usepackage{footnote}
\usepackage{lscape}
\usepackage{cancel}

\newcommand{\kms}{km\,${\rm s}^{-1}$}
\newcommand{\Msun}{M$_{\odot}$}
\newcommand{\west}{Cl$^{*}$\,Westerlund\,1\,W\,243}
\usepackage{graphicx}

\usepackage{txfonts}

\begin{document}

   \title{Multiplicity of Galactic Luminous Blue Variable stars
          \thanks{Based on observations collected at the ESO Paranal
          observatory under ESO program 0102.D-0460(A), on spectra obtained with the Mercator Telescope, operated on the island of La Palma by the Flemish Community, at the Spanish Observatorio del Roque de los Muchachos of the Instituto de Astrof{\'i}sica de Canarias, on observations made with the Southern African Large Telescope (SALT) under programme 2019-1-SCI-001, and with the TIGRE telescope, located at La Luz observatory, Mexico (TIGRE is a collaboration of the Hamburger Sternwarte, the Universities of Hamburg, Guanajuato, and Li{\`e}ge).}}

   \author{L.~Mahy\inst{1,2},
   C.~Lanthermann\inst{2,3},
   D.~Hutsem\'ekers\inst{4},
   J.~Kluska\inst{2},
   A.~Lobel\inst{1},
   R.~Manick\inst{5},
   B.~Miszalski\inst{6},
   M.~Reggiani\inst{2},
   H.~Sana\inst{2},
   E.~Gosset\inst{4}
   }

   \institute{Royal Observatory of Belgium, Avenue Circulaire 3, B-1180 Brussel, Belgium\\
   \email{laurent.mahy@oma.be}
   \and
   Institute of Astronomy, KU Leuven, Celestijnenlaan 200D, 3001, Leuven, Belgium
   \and
   The CHARA Array of Georgia State University, Mount Wilson Observatory, Mount Wilson, CA 91023, USA
   \and
   Research Director F.R.S.-FNRS, Space sciences, Technologies and Astrophysics Research (STAR) Institute, Universit{\'e} de Li{\`e}ge, All{\'e}e du 6 Ao{\^u}t, 19c, B{\^a}t B5c, 4000 Li{\`e}ge, Belgium
   \and
   South African Astronomical Observatory, PO Box 9, Observatory, 7935, South Africa
   \and
   Australian Astronomical Optics - Macquarie, Faculty of Science and Engineering, Macquarie University, North Ryde, NSW 2113, Australia
    }
   \date{Received xx Month Year; accepted xx Month Year}

  \abstract
   {Luminous Blue Variables (LBVs) are characterised by strong photometric and spectroscopic variability. They are thought to be in a transitory phase between O-type stars on the main-sequence and the Wolf-Rayet stage. Recent studies also evoked the possibility that they might be formed through binary interaction. Up to now, only a few are known in binary systems but their multiplicity fraction is still uncertain.}
   {This study aims at deriving the binary fraction among the Galactic LBV population. We combine multi-epoch spectroscopy and long-baseline interferometry to probe separations from 0.1 to 120 mas around confirmed and candidate LBVs. }
   {We use a cross-correlation technique to measure the radial velocities of these objects. We identify spectroscopic binaries through significant RV variability with an amplitude larger than 35\,\kms. We also investigate the observational biases to take them into account to establish the intrinsic binary fraction. We use \textsc{CANDID} to detect interferometric companions, derive their flux fractions, and their positions on the sky. }
   {From the multi-epoch spectroscopy, we derive an observed spectroscopic binary fraction of $26_{-10}^{+16}$\%. Considering period and mass ratio ranges from $\log(P_{\rm orb})= 0 - 3$ (i.e., from 1 to 1000 days), and $q=0.1 - 1.0$, respectively, and a representative set of orbital parameter distributions, we find a bias-corrected binary fraction of $62_{-24}^{+38}$\%. From the interferometric campaign, we detect 14 companions out of 18 objects, providing a binary fraction of 78\% at projected separations between 1 and 120 mas. From the derived primary diameters, and considering the distances of these objects, we measure for the first time the exact radii of Galactic LBVs to be between 100 and $650\,R_{\odot}$, making unlikely to have short-period systems among LBV-like stars. }
   {This analysis shows for the first time that the binary fraction among the Galactic LBV population is large. If they form through single-star evolution, it means that their orbit must be initially large. If they form through binary channel that implies that either massive stars in short binary systems must undergo a phase of fully non-conservative mass transfer to be able to sufficiently widen the orbit to form an LBV or that LBVs form through merging in initially binary or triple systems. Interferometric follow-up would provide the distributions of orbital parameters at more advanced stages and would serve to quantitatively test the binary evolution among massive stars.}
   \keywords{stars: variables: S Doradus - stars: evolution - binaries: general - stars: massive}

   \authorrunning{Mahy et al.}

   \maketitle
%

\section{Introduction}\label{Sec:intro}


Luminous Blue Variables are enigmatic objects located in the upper part of the Hertzsprung-Russell diagram (HRD). They have long been considered to be a brief evolutionary phase describing massive stars in transition to the Wolf-Rayet (WR) phase \citep{lamers02}. Although O and WR stars feature heavy mass-loss rates, the latter, when integrated over the stellar lifetime are usually not sufficient to enable a smooth direct evolution. At some point in their past, the progenitors of WR stars must have undergone a phase of extreme mass loss, during which their outer envelopes were removed to reveal the bare core that became the WR star. This extreme mass loss is thought to occur either during a Red Supergiant (RSG) or a Luminous Blue Variable (LBV) phase \citep{maeder00}.

However, this traditional view of LBVs as single-star evolutionary phase was strongly questioned \citep{gallagher89}. Theory \citep{groh13} and observations \citep[see e.g.][]{gal-yam07, kiewe12} showed that some progenitors of core-collapse supernovae (CCSNe, especially Type IIn) appear to be LBVs while theoretically the latest stage before exploding as CCSNe was expected to be either the Red Supergiant, Blue Supergiant, or WR stage. This significantly questions the hypothesis that LBVs are massive stars in transition to core He-burning objects. Recently, the relative isolation of LBVs was the subject of a lively debate \citep{smith15, smith16, humphreys16, davidson16, aghakhanloo17, aadland18, smith19b}. According to \citet{smith15}, when one compare the population of LBVs with those of O- or WR-type stars, the relative isolation of LBVs suggests that it is unlikely that their population turns into the observed population of WR stars. These authors suggested that LBVs are not concentrated in young massive clusters with early O-type stars as it would be expected from single-star evolution. Rather, LBVs would be largely products of binary or multiple systems that would have been kicked far from their birth places by explosion of their companions. This binary product scenario was already proposed by \citet{justham14} in which an early case B accretion (soon after the core H burning phases) could allow the massive star to gain sufficiently mass to reach core collapse with the properties expected for an LBV (spun-up, chemically enriched, and rejuvenated).

The evolutionary origin of LBVs is thus still an open question in modern astrophysics. What we know about LBVs is that they are massive, hot stars with very high mass-loss rates, located near the Eddington limit. The only way to classify a star as LBV is by studying its variability \citep{weis20}, either photometric or spectroscopic. To be classified as classical LBV, a star must undergo S-Doradus (hereafter S-Dor) cycles, i.e., temperature changes at practically constant luminosity \citep{humphreys94,groh09,smith17}, or giant eruptions. The S-Dor cycle consists of two different phases of variability: 1) a cool phase caused by its atmosphere that develops through optically thick winds, exhibiting spectra that look like A- or F-type stars, and 2) a quiescence phase during which the star has features similar to those of early B or O  supergiant or even WR stars. Up to now, the true mechanism that produces S-Dor cycles remains still uncertain even though several hypotheses have been developed to explain this variability, e.g., proximity of the Eddington limit, pulsations, binarity, turbulent pressure and sub-surface convection, wind-envelope interaction, etc (see \citealt{humphreys94} and references therein, \citealt{grafener12}, \citealt{sanyal15}, \citealt{grassitelli21}, among others).

Some stars also present common characteristics with the classical LBVs but have never been detected as showing S-Dor variability or giant eruption: the {\it dormant} or {\it candidate} LBVs \citep{humphreys94}. These stars are also located close to the Humphreys-Davidson limit in the upper part of the HRD. In the present paper, we refer to the confirmed and candidate LBVs under the general term: LBV-like objects.

The most studied star in this LBV-like class of object, $\eta$~Car, was suspected to be the result of a merging induced by three stars, kicking out the original primary star, and responsible for the great eruption observed in the 1800s \citep{portegieszwart16,smith18}. Since the detection of $\eta$~Car as a binary, other LBVs and LBV-like objects were identified as binary systems: HD~5980 \citep{koenigsberger10} in the Small Magellanic Cloud, HDE~269128 in the Large Magellanic Cloud, HR~Car \citep{boffin16}, HD\,326823 \citep{richardson11}, Pistol star \citep{martayan12} and MWC\,314 \citep{lobel13} all located in the Milky Way. Despite these discoveries, the binary fraction among the population of LBV-like objects has not yet been unveiled. Through their X-ray survey, \citet{naze12} detected 4 objects ($\eta$~Car, Schulte~12, GAL~026.47$+$00.02 and Cl$^{*}$~Westerlund~1~W~243) for which their X-ray detections are reminiscent of wind-wind collisions in binary systems, and 5 objects (P~Cyg, AG~Car, HD~160529, HD~316285 and Sher~25) where the X-ray detections reach a strong limit that can be due to binarity among other explanations. They concluded from half of the sample of Galactic LBV-like objects that the binary fraction must be between 26\% and 69\%. \citet{martayan16} investigated the environments that surround 7 LBV-like stars and searched for the presence of possible wide-orbit companions. From infrared images, they found that 2 out of 7 objects (HD~168625 and MWC~314) might be wide-orbit binaries, deducing from low-number statistics that about 30\% of LBV-like stars are in binary systems.

Assuming that the LBV phase is transitory between the O and WR stages, we would expect to have more objects in binary systems. It is indeed quite surprising that more than 70\% of the O-type stars \citep{sana12, sana14,moe18}, 40\% of the Be stars \citep{Oudmaijer10}, and over 50\% of the WRs \citep{vanderhucht01,dsilva20} are detected to be part of multiple systems, whilst the binarity among the LBV-like objects is expected to be approximately 30\%. The detection of gravitationally bound companions is difficult through spectroscopy given 1) the difference of luminosity between the components and 2) the strong variability characterising the LBV-like stars (as \citealt{weis20} mentioned in their review). In the present paper, we propose to combine spectroscopy and interferometry to study the multiplicity properties of a sample of Galactic confirmed and candidate LBVs using multi-epoch and multi-instrument observations. The paper is organised as follows. Section\,\ref{Sec:Obs} describes the sample, the multi-epoch and multi-instrument observations, and their reduction. The spectroscopic observations are analysed in Sect.\,\ref{Sec:RVmeas}. Section\,\ref{Sec:Astro} presents the interferometric detections. Finally, Sect.\,\ref{Sec:discussion} discusses the impact of the overall binary fraction on the evolutionary nature of the LBVs while the conclusions are given in Sect.\,\ref{Sec:conclusion}.

\section{Sample, observations and data reduction}\label{Sec:Obs}
\subsection{Sample}
We built our sample from the LBV catalogues of \citet{clark05} and \citet{naze12}. The spectroscopic and interferometric samples are different because the instrumental and observational constraints are different.

\begin{figure}[t!]\centering
    \includegraphics[width=9cm, trim=40 0 70 10,clip]{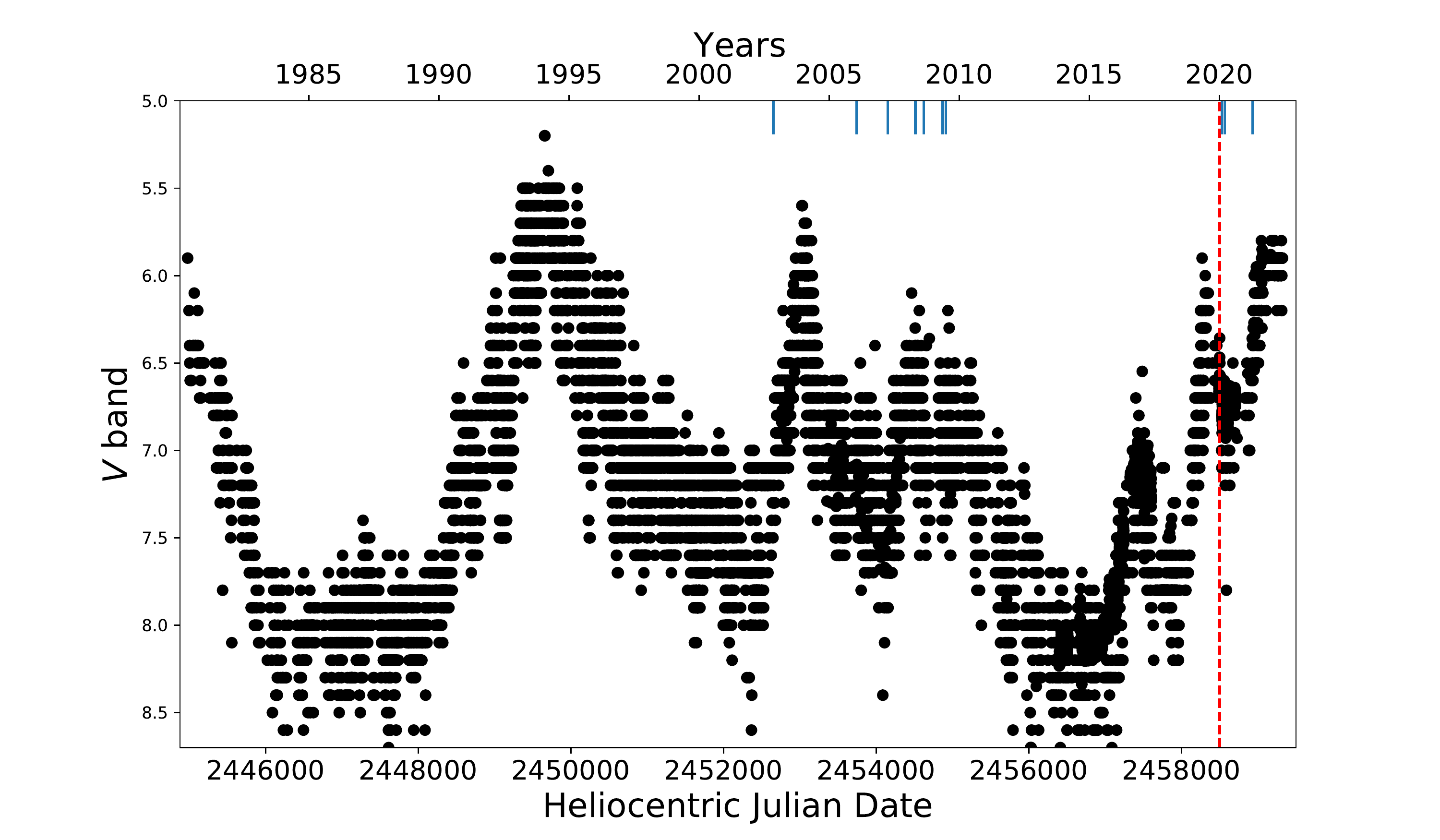}
     \includegraphics[width=9cm, trim=40 0 70 10,clip]{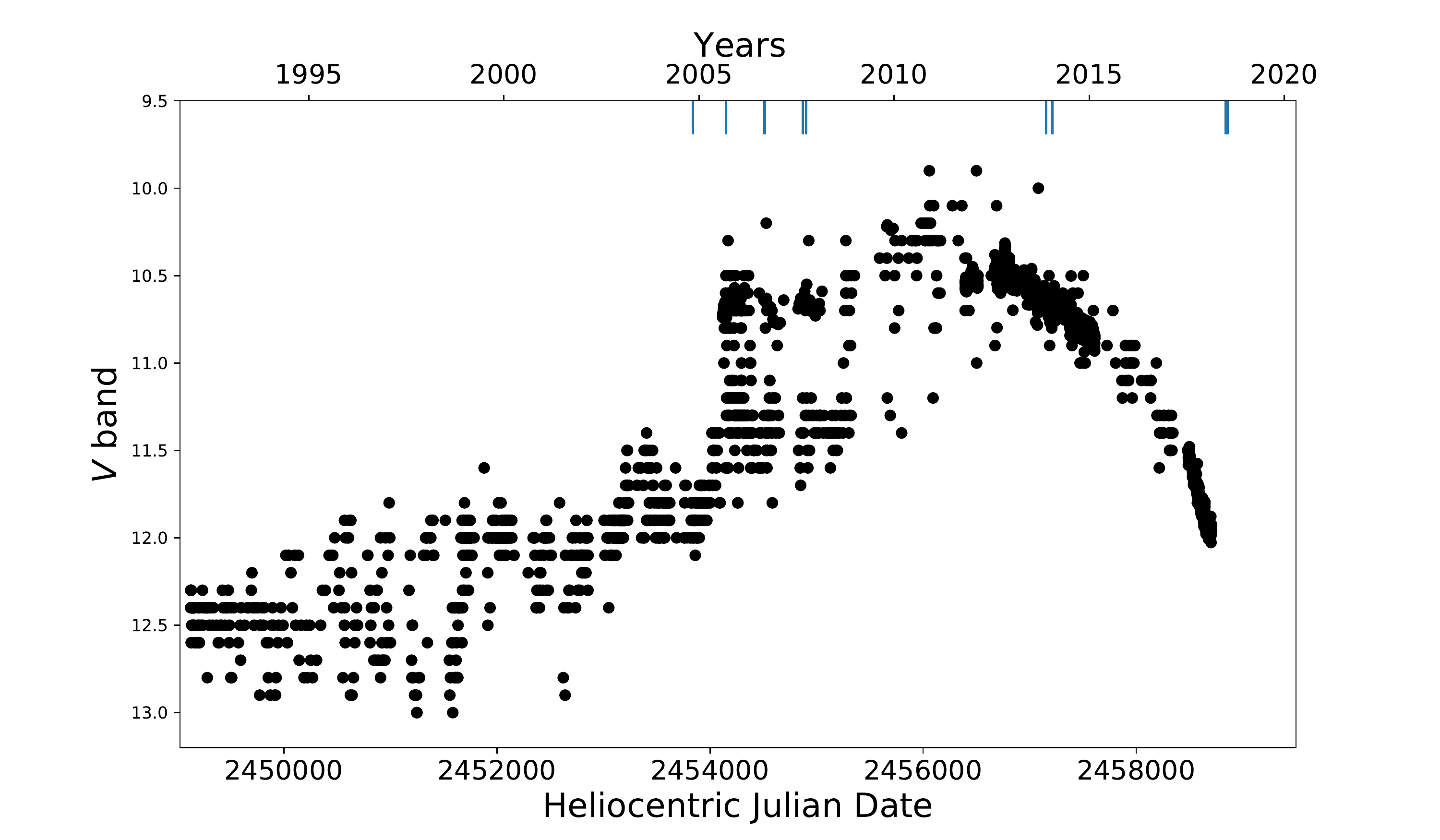}
    \caption{\label{Fig:AGCar_lc} AAVSO light curve of AG\,Car (top) and WRAY~15-751 (bottom) observed from Jan 1st 1982 to Apr. 20th 2021. The solid blue lines represent the Heliocentric Julian dates of the archival spectroscopy and the red dashed line shows the Heliocentric Julian date of the interferometric observation.}
\end{figure}

The spectroscopic sample gathers all the Galactic LBV-like stars with $V$ magnitudes brighter than $13.5$\,mag. We also included \west\ and Pistol star that are fainter than $V= 13.5$ because archival data exist. We remove AG\,Car because the spectra in the archives are too affected by the S-Dor cycle to measure with accuracy the RVs. We display in Fig.\,\ref{Fig:AGCar_lc} the American Association of Variable Star Observers (AAVSO) light curves\footnote{https://www.aavso.org.} of AG~Car (top panel) and  WRAY~15-751 (bottom panel) to show the photometric variability linked to their S-Dor cycles together with the spectroscopic observations. The light curve of AG~Car is clearly affected by a strong variability which prevents us to derive accurate radial velocities while for WRAY~15-751 we only discard the first and the last epochs from our spectroscopic dataset. We do not consider for our analysis $\eta$\,Car because this object was intensively analysed in the past few years and its detection as spectroscopic binary was already published by \citet{damineli96}. The final spectroscopic sample analysed in the present study thus contains 18 stars.

Because of the sensitivity of the instruments, the interferometric sample consists of stars brighter than $H= 5$\,mag in the North and $H=6.5$\,mag in the southern hemisphere. These stars must also be brighter than $13.5$\,mag in the $V$ band to be detected in the optical with 1-2m class telescopes, which prevents us to observe in interferometry the objects that are very extincted. Given these constraints, we came up with an interferometric sample of 15 stars, which we obtained data for (ESO Program ID: 0102.D-0460, PI: Mahy). To extend this sample, we also use the interferometric results published for $\eta$\,Car by \citet{gravity18}, HR\,Car by \citet{boffin16} and Pistol star by \citet{martayan12}, bringing the total of interferometric targets to 18 stars.

As mentioned above, both the spectroscopic and interferometric samples analysed in the present study contain 18 stars. Given the different requirements for the observations, only 11 targets belong to both samples. The list of the spectroscopic targets is given in Table\,\ref{table:spec} and that of the interferometric targets is given in Table\,\ref{table:interf}. We have marked with an asterisk the objects that belong to the two tables.

\subsection{Spectroscopy}\label{subsec:spectro}
For stars with declinations higher than $-25^{\circ}$, we collected observations with the High-Efficiency and high-Resolution Mercator Echelle Spectrograph (HERMES) mounted on the 1.2m Mercator telescope \citep{raskin11} at the Observatorio del Roque de los Muchachos in La Palma (Spain). The data were taken in the high-resolution fibre mode, which has a resolving power of $R = 85,000$, and covers the $4000-9000$\,\AA\ wavelength domain. The raw exposures were reduced using the dedicated HERMES pipeline and we worked with the extracted cosmic-removed, merged spectra afterwards.

Between 2013 and 2020 we have also obtained about 25 intermediate-resolution spectra of HD\,168625, HD\,168607 and P~Cygni, among other luminous stars, with the robotic 1.2m Telescopio Internacional de Guanajuato Rob{\'o}tico Espectrosc{\'o}pico \citep[TIGRE;][]{Schmitt2014} installed at the La Luz Observatory (Mexico). The TIGRE is equipped with the refurbished fibre-fed HEROS {\'e}chelle spectrograph which delivers spectra covering the wavelength ranges 3500$-$5600\AA\ (blue channel) and 5800$-$8800\AA\ (red channel) with a resolving power of about 20,000. Data reduction was done with the dedicated TIGRE/HEROS reduction pipeline \citep{Mittag2010}.

For stars in the southern hemisphere, we collected spectra with the High Resolution Spectrograph (HRS) on SALT \citep{bramall10,bramall12,crause14} under programme 2019-1-SCI-001 (PI: Miszalski). The data were taken in high-resolution mode for the brightest stars and in low-resolution mode for the faintest ones with the cut-off around $V=10$~mag. The data were reduced with the \textsc{midas} pipeline \citep{kniazev2016} based on the {\'e}chelle \citep{ballester92} and feros \citep{stahl99} packages. We applied heliocentric corrections to the data and checked the wavelength calibrations by using the Diffuse Interstellar Bands (DIBs) that are present within the wavelength coverage.

Finally, we retrieved spectra from the European Southern Observatory (ESO) archives observed with FEROS, UVES, HARPS and X-Shooter. FEROS \citep[Fibre-fed Extended Range Optical Spectrograph,][]{Kaufer99} is mounted on the MPG/ESO 2.2m telescope at La Silla (Chile). FEROS provides a resolving power of $R = 48,000$ and covers the entire optical range from $3800$ to $9200$\,\AA. The data were reduced following the procedure described in \citet{mahy10, mahy17}.

Some stars were observed with the UV and Visible Echelle Spectrograph \citep[UVES,][]{dekker00} mounted on the 8.2m Unit Telescope 2 (UT2)  telescope of the ESO Very Large Telescope (VLT) at the Paranal observatory (Chile). UVES has a resolving power of $R = 80,000$ and, depending on the setup, covers different wavelength ranges from the near-UV to optical domains. The spectra were reduced with the UVES pipeline data reduction software.

High-resolution, optical spectra were acquired with the High Accuracy Radial velocity Planet Searcher (HARPS, \citealt{mayor03}) spectrograph attached to the 3.6-m telescope at La Silla Observatory (Chile). In the high-efficiency (EGGS) mode, the spectral range covered is 3780-6910\,\AA\ and the resolving power $R \sim 80,000$. The data were reduced with the HARPS pipeline.

For very extincted objects, we also found archival spectra collected with X-Shooter \citep{vernet11} on the ESO/VLT UT2. X-Shooter is an intermediate resolution ($R \sim 4,000-17,000$) slit spectrograph covering a wavelength range from 3000 to $25,000$\,\AA, divided over three arms: UV-Blue (UVB), visible (VIS), and near-infrared (NIR). Since the objects are very extincted only the NIR arm was used to measure the radial velocities (RVs).

Unlike O-type stars on the main sequence, LBV-like spectra may display spectral lines with P-Cygni profiles or in emission. When they are in emission, those lines are however less broad than emission lines visible in WR spectra, which makes the definition of the continuum easier in LBV-like spectra. To define this continuum, we focus on wavelength regions free from P-Cygni profiles or emission lines. We reconstructed the continuum using splines of low order over limited wavelength windows. The observed spectra were then divided by the continuum to yield the normalised spectra.

\subsection{Interferometry}\label{subsec:interfero}

The bulk of the interferometric observations has been obtained at ESO (PI: Mahy, Program ID: 0102.D-0460) with the Very Large Telescope Interferometer (VLTI). All long baseline interferometric data were obtained with the Precision Integrated-Optics Near-infrared Imaging ExpeRiment (PIONIER) combiner \citep{lebouquin11,lebouquin12} and the four auxiliary telescopes of the VLTI. We used the ``large'' (A0-G1-J2-J3) and ``small'' (A0-B2-C1-D0) configurations with the auxiliary telescopes. PIONIER is a four-beam interferometric combiner in the near-infrared $H$-band (central wavelength of $1.65~\mu m$). The $H$-band is covered by 6 wavelength channels ($R \sim 40$). This instrument provides two observables: the squared visibilities (V2) that are related to the size of the target and closure phases (CP) that are related to the degree of (non-)point-symmetry.
Data were reduced and calibrated with the $pndrs$ package described in \citet{lebouquin11}. The statistical uncertainties typically range from $0.3^{\circ}$ to $12^{\circ}$ degrees for the CP and from 1 to 6\% for the V2, depending on target brightness and atmospheric conditions.

Each observation sequence of the targets in our sample was bracketted by two observations of calibration stars in order to master the instrumental and atmospheric response. These calibrators were found using SearchCal\footnote{\url{http://www.jmmc.fr/searchcal}} and selected to be close to the science object both in terms of position (within $\sim 2$ degrees) and magnitude (within $\pm1.5$~mag).

We also used the Michigan InfraRed Combiner-eXeter \citep[MIRC-X,][]{anugu20}, which is a highly sensitive six-telescope interferometric imager installed at the CHARA Array \citep[USA,][]{CHARA}. This instrument provides the V2 and CP observables as well. As this instrument combines the 6 telescopes of the CHARA array, a single observation with the 6 telescopes provides a sufficient (u,v)-coverage to detect and characterise a multiple system. MIRC-X is working in the $H$-band as well. The observations used in this paper were obtained using the PRISM50 mode (R $\sim$ 50). MIRC-X data were reduced using the public pipeline\footnote{\url{https://gitlab.chara.gsu.edu/lebouquj/mircx_pipeline}} described in \citet{anugu20}.

\section{Spectroscopic binary fraction}
\label{Sec:RVmeas}

\subsection{Radial velocities and the multiplicity criteria}\label{Subsec:RVmeas}

\subsubsection{RV measurements}\label{Subsec:rvmeas}

In order to assess the RV variability of each star in a systematic way, we measure the Doppler shifts of a set of spectral lines in all available epochs (as long as the spectra are not affected by S-Dor cycles). The list of the selected spectral lines is given in Table\,\ref{table:lines}. In addition, we took a special care to only select spectra that have been collected during a same hot quiescent or cool eruptive phase, given the spectral variability produced through the S-Dor cycles. These cycles have typical timescales of about a decade which allows us to probe long-period (longer than a few years) spectroscopic binary systems. The individual properties of the different campaigns are given in Table\,\ref{table:spec}

\begin{figure*}[t!]\centering
    \includegraphics[width=19cm, trim=80 30 70 70,clip]{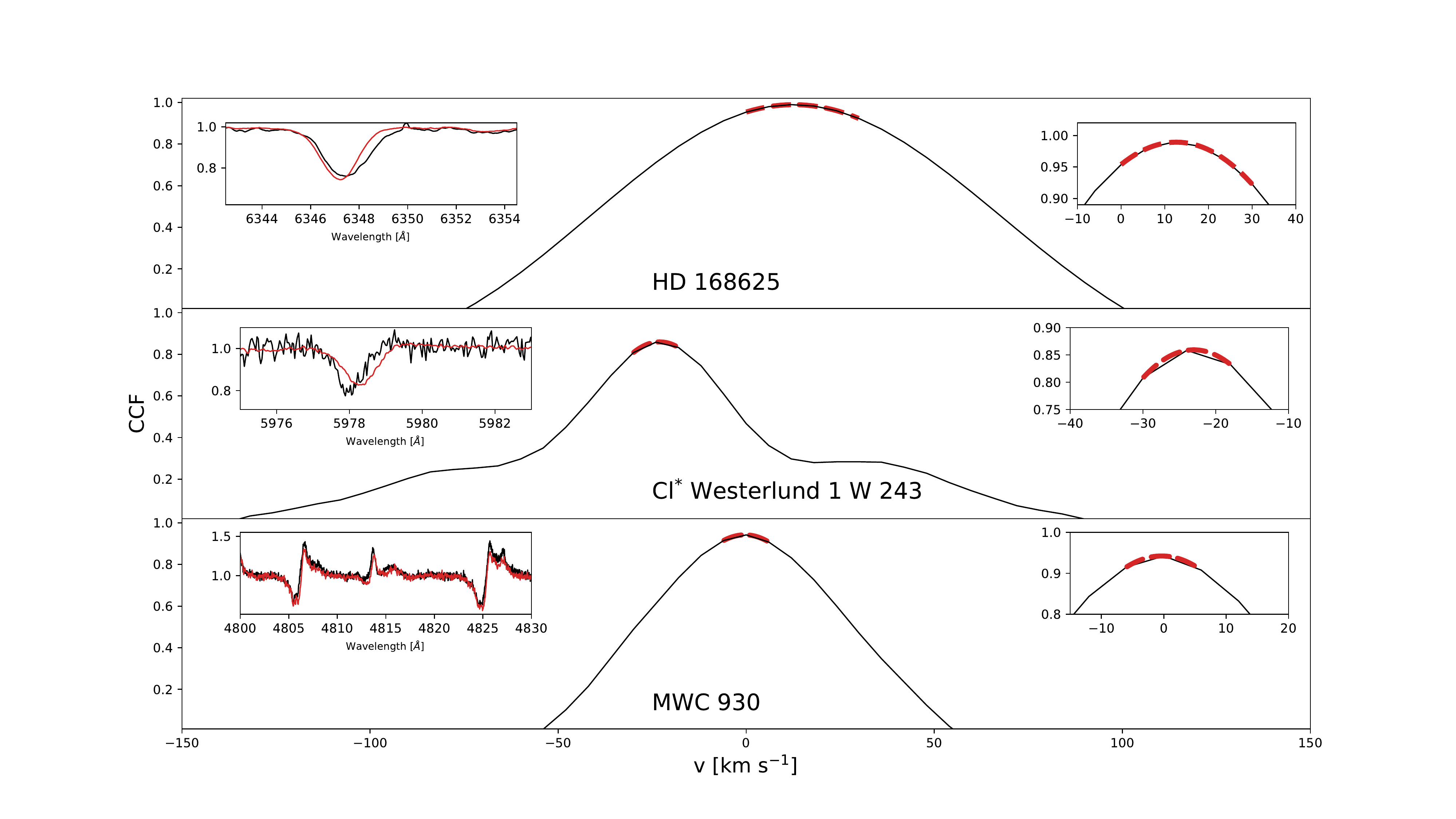}
    \caption{\label{Fig:CCF} Examples of CCF for three stars in our sample: HD~168625 (top), \west\ (middle) and MWC~930 (bottom). A zoom-in is displayed on the left-hand side of the figure to show the comparison between the template (red) and one of the observations (black). The right-hand side onsets show zoom-in on the top part of the CCF and in red the parabola fit to derive the RVs.}
\end{figure*}

Before measuring the RVs, we carefully inspected the line profiles to look for clear signatures of putative companions that could move in anti-phase and that can be directly related to binarity. We did not find any spectral features that can be obviously related to a spectroscopic companion. We measured the RVs of the LBV-like stars by cross-correlating specific spectral lines or a whole region of lines with a template and then fit a parabola to the maximum region of the cross-correlation function (CCF, see \citealt{shenar17}). Figure\,\ref{Fig:CCF} displays the CCF for three stars in our sample with the parabola fit of the maximum region. The choice of the lines/regions to cross-correlate with depends on the target and on the wavelength coverage. Since the spectroscopic data have been collected with different instruments, their wavelength coverage is not uniform between the whole dataset. We thus focus on spectral lines that are common to all the spectra (Table\,\ref{table:lines}). We mainly favour the selection of absorption lines because they are formed closer to the photosphere of the stars. Several objects however only exhibit P-Cygni profiles or emission lines as it is the case for WR31a, or P~Cyg for instance. 

The cross-correlation technique that we use in the present work is based on \citet{zucker03}. Given the complexity of modelling LBV-like spectra with an atmosphere modelling code (beyond the scope of this paper), the templates used for the cross-correlation have been chosen initially to be one of the observations, i.e., the spectrum with the highest signal-to-noise ratio (SNR). One of the advantages of using an observation as a template is that it is not affected by the fact that different spectral lines of LBV-like stars may imply different RVs due to their varying formation regions and asymmetric profiles. The spectra are then co-added in the rest-frame of the star using these RVs to create a higher SNR template, which is then used to iterate on the RV measurements. We generally did three iterations before obtaining the final sets of RVs. As done by \citet{shenar17}, the absolute RVs are obtained by cross-correlating the high SNR template with a suitable atmosphere model. Table\,\ref{table:radvel} gives the RV measurements for each spectral line/region and the mean values computed through all the lines/regions used for a given star. We also indicate the spectrum that was initially used as template for the cross-correlation. A full version of this table is available from the CDS.

\begin{table*}
\caption{Spectral lines used for cross-correlation}              
\label{table:lines}      
\centering                                      
\begin{tabular}{l l}          
\hline\hline                        
Star & Rest wavelength\\
\hline
AS\,314 & \ion{Si}{ii}~4128-4130~/~\ion{He}{i}~4471~/~\ion{Mg}{ii}~4481~/~\ion{Si}{ii}~6347-71\\
HD\,160529 & \ion{Si}{ii}~4128-4130~/~\ion{Mg}{ii}~4481\\
HD\,168607 & \ion{Si}{ii}~4128-4130~/~\ion{He}{i}~4471~/~\ion{Mg}{ii}~4481~/~\ion{Si}{ii}~6347\\
HD\,168625 & \ion{He}{i}~5876~/~\ion{Si}{ii}~6347-71~/~\ion{C}{ii}~6578-83\\
HD\,316285 & \ion{He}{i}~4471~/~\ion{Mg}{ii}~4481~/~\ion{Fe}{ii}~4526-34-36~/~\ion{N}{ii}~5666-76-79-86~/~\ion{Al}{III}~5723~/~\ion{S}{ii}~6449\\
HD\,326823 & \ion{N}{ii}~5666-76-79-86 \\
MWC~314 & \ion{S}{ii}~5454-74~/~\ion{Ne}{i}~6402\\
MWC~930 & \ion{He}{i}~4471~/~\ion{Mg}{ii}~4481~/~\ion{N}{ii}~4630~/~\ion{Si}{iii}~4813-19-28\\
WR31a & \ion{He}{i}~4471~/~\ion{Mg}{ii}~4481~/~\ion{N}{ii}~4601-07-14-21-30~/~\ion{N}{ii}~5666-76-79-86 \\
HD\,80077 & \ion{He}{i}~4471~/~\ion{Mg}{ii}~4481~/~\ion{Si}{iii}~4552~/~\ion{N}{ii}~4601-07-14-21-30~/~\ion{Fe}{iii}~5156~/~\ion{N}{ii}~5666-76-79-86 \\
WRAY~15-751  & \ion{Mg}{ii}~4481~/~\ion{Si}{ii}~6347 \\
HR\,Car  & \ion{N}{ii}~5666-76-79-86~/~\ion{Al}{III}~5723 \\
P~Cyg & \ion{S}{iii}~4253~/~\ion{Si}{iii}~4552~/~\ion{N}{ii}~4601-07-14-21-30~/~\ion{N}{ii}~5666-76-79-86\\
Sher~25  &  \ion{He}{i}~4471~/~\ion{Mg}{ii}~4481~/~\ion{N}{ii}~5666-76-79-86~/~\ion{He}{i}~ 5876\\
$\zeta$\,Sco & \ion{Si}{iii}~4813-19-28~/~\ion{N}{ii}~5666-76-79-86 \\
Pistol & \ion{Mg}{ii}~21,347-21,438 \\
Schulte~12 & \ion{N}{ii}~5666-76-79-86~/~\ion{He}{i}~5876~/~\ion{Si}{ii}~6347\\
\west & \ion{Si}{ii}~5979~/~\ion{Fe}{ii}~6179~/~\ion{Si}{ii}~6347-71 \\
\hline                                             
\end{tabular}
\end{table*}

\subsubsection{Multiplicity criteria}
\label{subsec:multiCrit}
To classify which LBV-like stars can be assessed as binary candidates, we have to determine whether the measured RV variability is statistically significant or not. We consider two statistical multiplicity criteria described in \citet{sana13}.
A star is classified as a likely spectroscopic binary if at least one pair of RVs measured at different epochs satisfies simultaneously these criteria:
\begin{enumerate}
    \item $\frac{|v_i - v_j|}{\sqrt{\sigma_i^2 + \sigma_j^2}} > 4.0$,
    \item $\Delta {\rm RV} = |v_i - v_j| > \Delta {\rm RV_{min}}$.
\end{enumerate}
where $v_i$ and $v_j$ are the individual RV measurements and $\sigma_i$ and $\sigma_j$ the respective $1\sigma$ errors on the RV measurements at epochs $i$ and $j$.
The first criterion defines a statistical test under which the null hypothesis of constant RV is rejected if, for a given star, any two RV measurements deviate significantly from one another. This criterion searches for significant variability at a $4\sigma$ level \citep{sana13}, considering the uncertainties on the measurements.
The second criterion considers that the observed variability detected through the RV measurements is actually caused by orbital motion in a binary system and not through intrinsic variability that can be induced by pulsations, atmospheric activity or, in the case of LBV-like stars, by wind effects. We therefore adopt that two individual RV measurements have to deviate from one another by more than a minimum amplitude threshold $\Delta {\rm RV_{min}}$.

\begin{table*}
\footnotesize
\caption{Sample of LBV-like stars with spectroscopic detection parameters. The errors correspond to $\pm 1\sigma$. The $^{*}$ symbol marks the objects that are also in the interferometric sample.}              
\label{table:spec}      
\centering                                      
\begin{tabular}{l r | c c r r r r}          
\hline\hline  
Star & Classif. & \# Spec. & Time cov. & Instru.\tablefootmark{a} & $\Delta$\,RV$_{\rm max}$  & $\sigma_{\rm RV}$ & Spec. bin \\
  & & & [d]& & [\kms] & [\kms] &   \\ 
\hline                                   
AS\,314 & cLBV & 9 & 1522.6 & F/H & 22.7 & 1.3 & N \\
HD\,160529$^{*}$& LBV & 6 & 5306.6 & F/U & 7.0 & 1.0 & N  \\
HD\,168607$^{*}$& LBV& 64 & 6931.9 & F/U/T/H & 27.8 & 1.9 & N \\
HD\,168625$^{*}$& cLBV& 60 & 7027.7 & F/U/T/H & 19.5 & 1.1 & N \\
HD\,316285$^{*}$& cLBV& 5 & 2656.8 & F/U & 20.6 & 2.2 & N \\
HD\,326823$^{*}$& cLBV & 8 & 5248.8 & F & 382.4 & 3.5 & Y \\
HD\,80077$^{*}$& cLBV & 11 & 2093.3 & F/S/U/X & 23.1 & 0.7 & N \\
HR\,Car$^{*}$ & LBV & 21 & 5396.3 & F/U/Ha & 58.5 & 0.7 & Y  \\
MWC\,314 & cLBV & 30 & 4040.0 & H & 164.4 & 2.9 & Y \\
MWC\,930& LBV & 16 & 5451.0 & F/U/H & 32.9 & 3.0 & N  \\
P\,Cyg$^{*}$& LBV & 46 & 4067.9 & H/T & 26.3 & 1.9 & N \\
Pistol & cLBV & 14 & 1439.9 & X & 11.9 & 0.1 & N \\
Schulte~12$^{*}$& cLBV & 15 & 3657.9 & H & 24.2 & 1.6 & N \\
Sher\,25& cLBV & 11 & 1723.1 & F & 19.3 & 1.7 & N  \\
W\,243$^{*}$& LBV & 6 & 373.0 & U & 30.1 & 1.9 & N  \\
WR\,31a & cLBV & 14 & 6538.3 & F/U/S & 20.3 & 1.3 & N \\
WRAY\,15-751& LBV & 12 & 5019.5 & F/Ha/S & 65.6 & 3.4 & Y \\
$\zeta$\,Sco$^{*}$& cLBV & 7 & 2612.7 & F/U/Ha & 13.6 & 0.5 & N \\
$\eta$ Car$^{*}$& LBV & -- & -- & -- & -- & -- &  Y \\
\hline                                        
\end{tabular}
\tablefoot{
\tablefoottext{a}{H: HERMES -- F: FEROS -- U: UVES -- X: XSHOOTER -- Ha: HARPS -- T: TIGRE -- S: SALT HRS}
}
\end{table*}

Previous studies about massive OB-type stars have used a variability threshold of 20\,\kms\ \citep[][Banyard et al. 2021 subm.]{sana12,sana13,bodensteiner21} to probe the multiplicity property in different metallicity environments (Galaxy, Large and Small Magellanic Clouds). The stars in these different samples are however massive stars on the main sequence with little or undetectable line profile variability that could impact their multiplicity status. To make the difference between intrinsic variability and orbital motion among B-type stars in the LMC, \citet{dunstall15} adopted the threshold value of 16\,\kms\ from their B-supergiant sample and used it over the entire population of B-type stars in the 30\,Dor region. Simon-Diaz et al. (2021, submitted) investigated the impact of pulsations on the multiplicity properties in a large population of Galactic OB supergiants. They collected spectra on 56 Galactic OB supergiants, 13 O dwarfs and subgiants and 5 early-B giants. They showed that the peak-to-peak amplitudes of RV measurements could reach up to 20-25\,\kms\ for late-O and early-B supergiants and decrease to between 2 and 15\,\kms\ for the O dwarfs and the late-B supergiants. When we look at more evolved stars such as the Wolf-Rayet (WR), \citet{dsilva20} observed two distinct kinks in their observed binary fraction computed over a sample of 12 objects classified as WC (carbon-rich WR) stars. The first one is detected in the ranges between 5 and 10\,\kms\ while the second is between 12 and 19\,\kms, leading to an observed spectroscopic binary fraction of 0.58 and 0.33, respectively. These observed spectroscopic binary fractions, once corrected for the observational biases, lead to an intrinsic binary fraction higher than 0.72. 

\begin{figure}[t!]\centering
    \includegraphics[width=9cm,trim=0 0 0 0,clip]{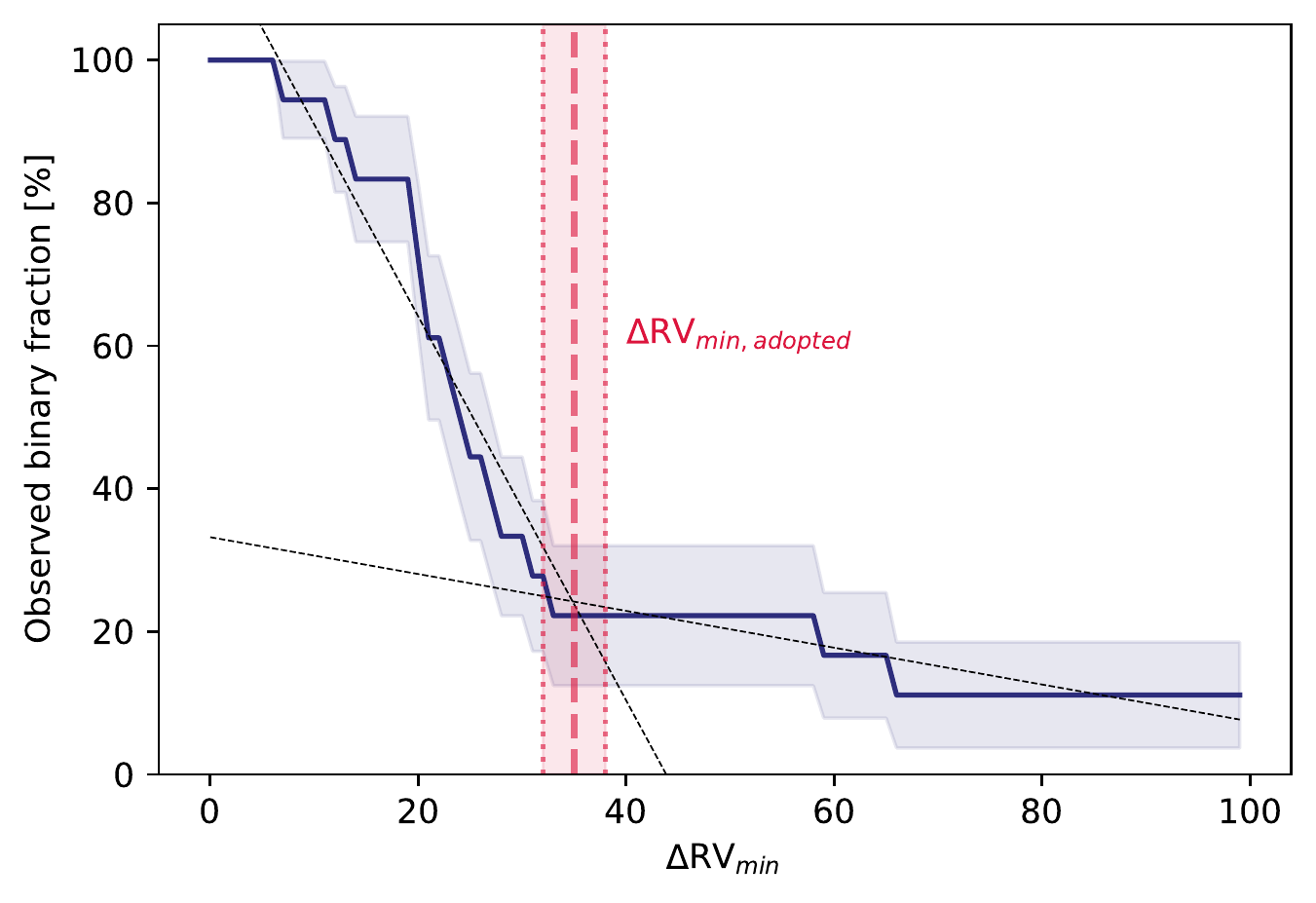}
    \caption{\label{Fig:binaryFrac} Observed binary fraction as a function of the adopted RV-variability threshold $C$. The vertical dotted line gives the adopted threshold value $\Delta {\rm RV_{min}}=35 \pm 3$\,\kms\ is indicated by the red line. The shaded area represents the $\pm 1\sigma$ error.}
\end{figure}

Adopting different values as threshold obviously leads to different observed binary fractions. However, when one corrects for observational biases, the choice of this threshold is taken into account, so that it does not affect the final binary fraction as long as a significant contamination by false positive is avoided. We investigate the impact of the choice of this threshold $\Delta {\rm RV_{min}}$ on the observed spectroscopic binary fraction among the population of LBV-like stars. We vary this threshold between 0 and 100\,\kms\ and calculate the observed spectroscopic binary fraction and the corresponding binomial error for each value of this threshold (Fig.\,\ref{Fig:binaryFrac}). We performed a linear regression on the 0-35\,\kms\ range and another linear regression on the 40-100\,\kms\ range of the distribution. The kink, showed by the different slopes in these two regressions, is at $\Delta {\rm RV_{min}} = 35 \pm 3$~\kms. We use this value as a threshold, which provides us with an observed spectroscopic binary fraction of $f_{\rm obs} = 0.22_{-0.10}^{+0.16}$, where binomial statistics has been used to compute the uncertainty. 

Our threshold is larger than those used for OB stars on the main sequence, for B supergiant populations, or for carbon-rich WR stars. As mentioned above, this choice has been dictated to consider the intrinsic variability of those object but has as consequences that possible binary systems might have not been detected. Given that our sample consists of confirmed and dormant LBVs, the choice of this threshold however appears as a good compromise to discriminate against processes that lead to RV variability and that are not linked to binarity (e.g., pulsations, winds, S-Dor cycles for the confirmed LBVs). We detect 4 binary candidates out of 18 objects in our sample. Among the detected candidates, there is HR\,Car that was already detected in interferometry with an orbital period of about 2330 days \citep[][Boffin et al. 2021 in prep.]{boffin16}. MWC\,314 and HD\,326823 were detected, by \citet{lobel13} and \citet{richardson11} through spectroscopy, as binary systems with orbital periods of about 6 and 61 days, respectively. Another system is identified as candidate binary: WRAY\,15-751. Two other objects lie just below the threshold of 35\,\kms: MWC~930 and \west. Both objects are confirmed LBVs. MWC~930 shows spectral variability and its spectral lines are broad and show splitting \citep{Miroshnichenko05,Miroshnichenko14}. \citet{lobel17} attributed this line splitting to optically thick central line emission produced in the inner ionized wind region. These authors suggested that this split is quite similar to the split observed in metal line cores of pulsating Yellow Hypergiants such as, e.g., $\rho$~Cas. \west\ was analysed by \citep[][]{ritchie09} who claimed that this object might be a binary system with an undetected hot OB companion. However, the RV measurements were not obviously consistent with binarity, except if the system is seen under a low inclination, or has a long period or is highly eccentric. For Schulte~12, the threshold that we adopt does not select that star as potential binary system. A period of 108 days has been reported from spectroscopy and X-rays by \citet{naze19}. This points out that this period might be associated with another phenomenon than binarity as suggested by these authors. We also fail to detect RV variability linked to binarity for HD\,168625 or Pistol star while those two objects were possible wide-orbit binary systems \citep{martayan16}. Additionally, the observed spectroscopic binary fraction rises up to $f_{\rm obs} = 0.26_{-0.10}^{+0.16}$ if we include the spectroscopic binary $\eta$~Car to the sample \citep{damineli96}. 

The time coverages of our data are varying from one object to another, since our spectroscopic data set is quite heterogeneous, but they allow us to probe time scales between 373 days for \west\ and about 7000 days for HD\,168607 and HD\,168625. Besides confirming the orbital periods found for MWC\,314 ($P_{\rm orb} \sim $ 60 days) and HD\,326823 ($P_{\rm orb} \sim $~6 days), no clear periodic signal has been found from our RV measurements. This suggests that either most of the LBV-like objects that pass our binary criteria are long-period systems or that the threshold is not strict enough. In any case, binary systems with orbital periods shorter than half the time coverage of the stars (see Table\,\ref{table:spec}) would have been detected from our spectra, except if they were systems seen under a low inclination, with a low mass ratio, or with a high eccentricity. 
 More details about the detection probability are given in Sect.\,\ref{subsec:binarydec}.

\subsection{Correction for observational biases}
\label{subsec:binarydec}
\begin{figure*}[t!]\centering
    \includegraphics[width=6cm,trim=10 10 10 10,clip]{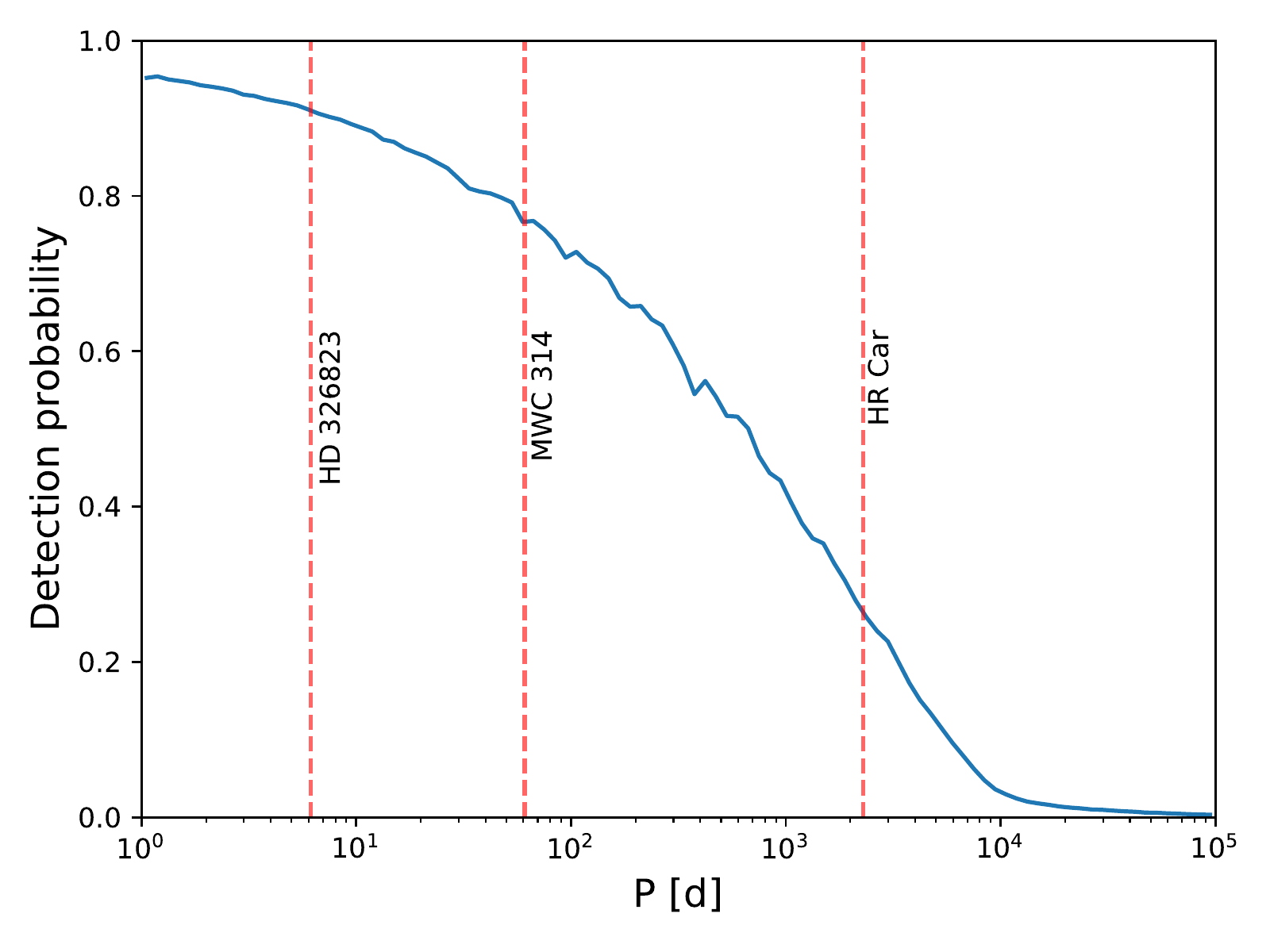}
    \includegraphics[width=6cm,trim=10 10 10 10,clip]{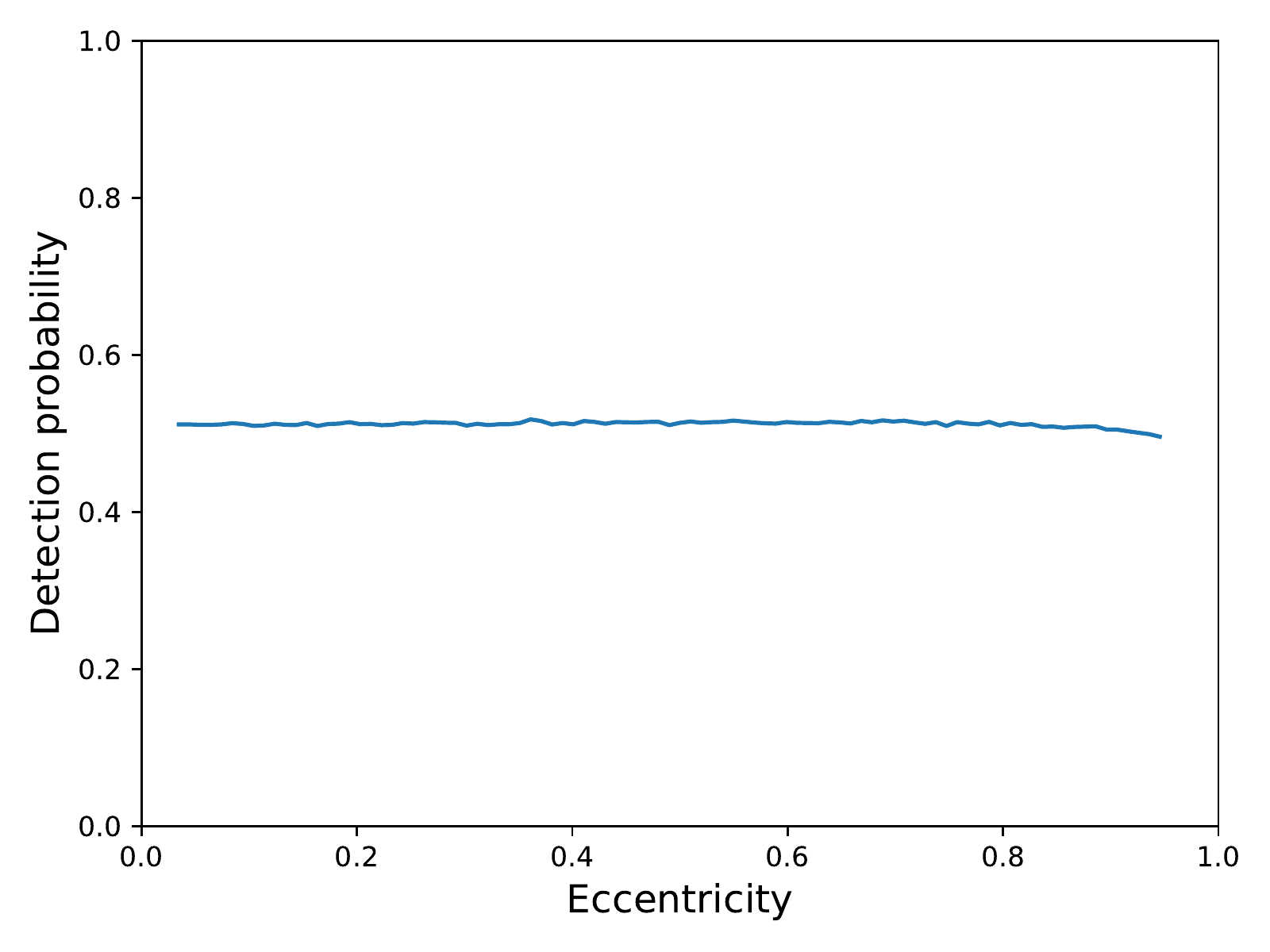}
    \includegraphics[width=6cm,trim=10 10 10 10,clip]{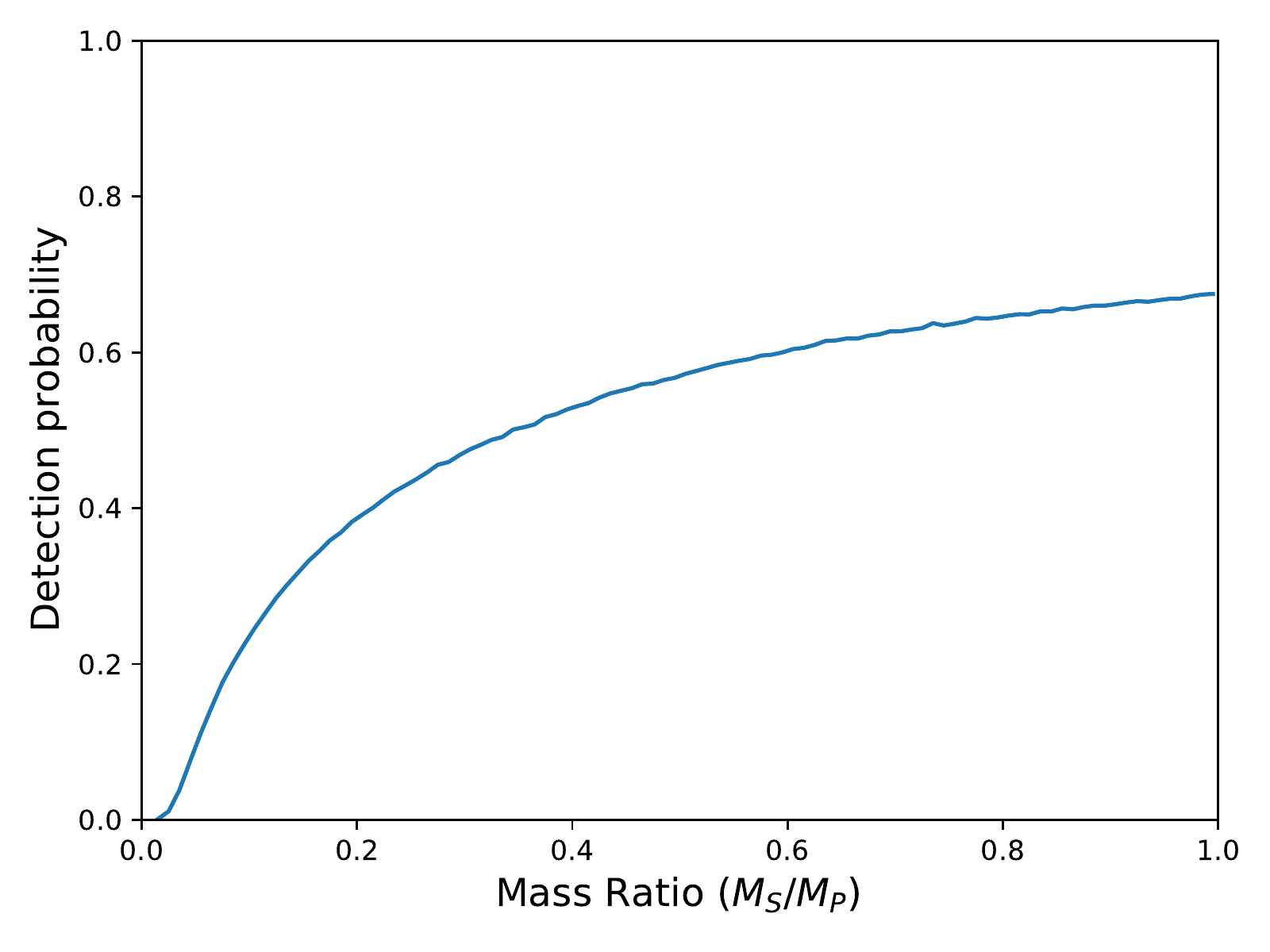}
    \caption{\label{Fig:detectionProb} Mean binary detection probabilities as a function of the orbital period (left), eccentricity (middle) and mass ratio (right). Known orbital periods of LBV-like stars in our sample are indicated by vertical lines. }
\end{figure*}

In order to assess the intrinsic multiplicity properties of the LBV-like objects, the observed binary fraction needs to be corrected for the observational biases. From our simulations and following the approach described in \citet{sana13}, the probability of detecting our objects as binaries (Fig.\,\ref{Fig:detectionProb}) gives us an estimate of the sensitivity of our observations, and provides us with a correction factor. The latter is multiplied by the observed spectroscopic binary fraction to estimate the intrinsic binary fraction. 

The first observational bias that we need to deal with is that the more spectra we get the better we can assert the binarity detection through spectroscopy. To quantitatively assess the sensitivity of our data to the parameter space as well as to the properties of the undetected binary systems, we computed the binary detection probabilities.

We investigate the orbital configurations (period, eccentricity, mass ratio) by assuming a mass range for the LBV-like objects that would yield to RV signals that are compatible with our observational constraints. We run 1 million of Monte Carlo simulations tuned for each target of our sample. The period distribution was taken uniform over the logarithmic space and ranges from 1 to $10^5$ days, the eccentricity between 0 and 0.95, with all orbits with eccentricities lower than 0.03 considered as circular. The mass-ratio distribution is considered uniform and ranges from 0.01 to 1.0. Finally, the initial masses are taken for each star from \citet{smith19} from their position in the Hertzsprung-Russell diagram. We adopt a Salpeter initial mass function (IMF, with $\alpha = -2.35$) for the primary masses with a tolerance varying from 5 to 40\,\Msun. We however emphasise that the derived fractions are not sensitive to the slope of the IMF because the considered mass range is fairly limited. Our simulations assume that the orbits are randomly oriented in the three-dimensional space, the time of periastron passage is uncorrelated with respect to the start of the RV campaign, and that the orbital parameters are uncorrelated. To detect one system as binary, it needs to fulfill the same criteria as we adopt for our analysis (Sect.\,\ref{subsec:multiCrit}).

Figure\,\ref{Fig:detectionProb} shows the average detection probability curves of our spectroscopic campaign computed from all the objects in our sample. Globally, the orbital periods shorter than $10^2$ days are well-covered and a detection probability close to 75\% is reached. If the binary systems have orbital periods within the range $10^2-10^3$ days, the detection probability drops between  $42$ and $73$\%. For extremely long-period systems (between $10^3$ and $10^4$ days), the detection probabilities fall from 42\% to 3\%, showing the difficulty to constrain them from spectroscopy and the need for other observational techniques such as interferometry or high-contrast imaging.

Given that our campaign is sensitive to orbital periods up to $10^3$ days, the overall detection probability over that period is about 42\%. Adopting different parameter distributions would however provide different values. As mentioned by \citet{bodensteiner21}, this impacts the intrinsic binary fraction within an error of about 5\%. While the overall observed spectroscopic binary fraction is estimated to $f_{\rm obs} = 0.26_{-0.10}^{+0.16}$ (including $\eta$~Car, see Sect.\,\ref{subsec:multiCrit}), the intrinsic, bias-corrected binary fraction is therefore computed to be $f_{\rm intrinsic} = 0.62_{-0.24}^{+0.38}$ for the Galactic LBV-like population.

\section{Interferometric companion detection}
\label{Sec:Astro}

\subsection{Multiplicity fraction and number of companions}

To search for astrometric companions among LBV-like stars, we used the \textsc{CANDID} code\footnote{\url{https://github.com/amerand/CANDID}} \citep{gallenne15}. \textsc{CANDID} is a set of \textsc{PYTHON} tools that was specifically created to search systematically for high-contrast companions. CANDID provides the best fit from a Levenberg-Marquardt minimisation that performs a systematic exploration of the parameter space. The best fit is found assuming a disc for the primary, an unresolved secondary, and some background flux. A disc used to mimic the primary can correspond to stellar photosphere, or to any material around the stars that are too close to the stellar surface to be resolved by interferometry (wind, dust torus,...). The reliability of the fit is only based on whether or not the grid to search for a companion is fine enough \citep{gallenne15}. All the detections performed by \textsc{CANDID} are given in Table\,\ref{table:interf}.

Among our sample, we detect 11 companions out of 15 LBV-like objects with a high number of sigmas ($>7\sigma$). For two of these companions, the best-fit models provided by \textsc{CANDID} fail to provide all the parameters even though their fits are reliable (HD\,168625 and Schulte\,12). For HD\,168625, the \textsc{CANDID} best-fit model did not allow us to fit the primary diameter, which tends to indicate that the primary star is unresolved. For Schulte\,12, the separation and position angle could not be derived. The best-fit provided by \textsc{CANDID} is reliable but the detected companion has a separation larger than 100\,mas from the central star, i.e. outside the interferometric field-of-view of MIRC-X with the current resolution.

\begin{table*}
\footnotesize
\caption{Sample of LBV-like stars with interferometric detection parameters. The errors correspond to $\pm 1\sigma$. The $^{*}$ symbol marks the objects that are also in the spectroscopic sample.}              
\label{table:interf}      
\centering                                      
\begin{tabular}{l r r r r r r r r r r}          
\hline\hline  
Star & Classif. & Instrum. & Prim. diam. & Flux frac. & $\Delta {\rm mag}$ & $\rho$ & PA & $n\sigma_{\rm det.}$ & $\chi^2_{\rm red}$ & remarks\\
& & & [mas] & [\%] & [mag] & [mas] & [$^{\circ}$] & & & \\ 
\hline                                   
HD\,160529$^{*}$& LBV & PIONIER & $1.13_{-0.01}^{+0.01}$ & $0.37_{-0.04}^{+0.02} $ & $6.1_{-0.1}^{+0.1}$ & $6.32_{-0.14}^{+0.24}$ & $328.36_{-1.05}^{+1.53}$ & 18 & 0.66  \\
HD\,168607$^{*}$& LBV& PIONIER &$0.93_{-0.01}^{+0.01}$ & $0.94_{-0.05}^{+0.03}$ & $5.1_{-0.1}^{+0.1}$& $21.66_{-0.19}^{+0.24}$ & $7.22_{-0.16}^{+0.24}$ & 7 & 0.43 \\
HD\,168625$^{*}$& cLBV &PIONIER & -- & $0.23_{-0.02}^{+0.23}$ & $6.6_{-1.0}^{+0.1}$ & $24.24_{-0.30}^{+0.30}$ & $299.32_{-0.84}^{+0.84}$ & 21 & 0.10 & no fitting diam. \\
HD\,316285$^{*}$& cLBV&PIONIER & -- & -- & -- & -- & -- & -- & -- & no detection\\
HD\,326823$^{*}$& cLBV &PIONIER & $0.88_{-0.03}^{+0.03}$ & $2.81_{-0.41}^{+0.41}$ & $3.9_{-0.2}^{+0.2}$ &  $1.43_{-0.13}^{+0.13}$ & $187.69_{-1.62}^{+1.62}$ & 24 & 3.95 \\
HD\,80077$^{*}$& cLBV &PIONIER & -- & -- & -- & -- & -- &-- & -- & no detection\\
HR\,Car$^{*}$ & LBV &PIONIER & $0.37_{-0.05}^{+0.07}$ & $9.9_{-0.5}^{+0.4}$ & $6.1_{-0.1}^{+0.1}$ & $1.41$ & $6.17$ & -- & 0.65 & \citet{boffin16} \\
P\,Cyg$^{*}$& LBV & MIRC-X & $0.76_{-0.01}^{+0.01}$ & $1.91_{-0.33}^{+0.32}$ & $4.3_{-0.2}^{+0.2}$&  $13.00_{-0.07}^{+0.11}$ & $131.75_{-0.28}^{+0.46}$ & 8 & 1.95 \\
Pistol & cLBV & PIONIER &-- & -- & -- & $\sim 65$ & $\sim 293$ & -- & --  & \citet{martayan11} \\
Schulte~12\tablefootmark{a$^{*}$}& cLBV & MIRC-X & $1.27_{-0.01}^{+0.01}$ & $14.00_{-1.40}^{+1.70}$ & $2.1_{-0.1}^{+0.1}$ & $ \sim 100$ & -- & 50 & 17.16 & more obs. needed\\
W\,243$^{*}$& LBV &PIONIER & $0.81_{-0.06}^{+0.07}$ & $5.50_{-0.42}^{+0.41}$& $3.1_{-0.1}^{+0.1}$& $1.51_{-0.04}^{+0.09}$ & $222.07_{-1.33}^{+3.22}$ & 50 & 1.45 \\
$\zeta$\,Sco$^{*}$& cLBV &PIONIER & $0.75_{-0.01}^{+0.01}$ & $0.30_{-0.03}^{+0.04}$ & $6.3_{-0.1}^{+0.1}$& $11.54_{-0.10}^{+0.10}$ & $283.22_{-0.34}^{+0.76}$ & 31 & 0.20 \\
AG\,Car& LBV &PIONIER & $1.08_{-0.01}^{+0.02}$ & $0.50_{-0.03}^{+0.07}$ & $5.8_{-0.1}^{+0.2}$ & $4.92_{-0.12}^{+0.22}$ & $100.19_{-1.28}^{+3.36}$ &7 & 1.48  \\
MN\,46$^{\dagger}$& cLBV &PIONIER & $1.40_{-0.02}^{+0.02}$ & $3.35_{-0.24}^{+0.34}$ & $3.7_{-0.1}^{+0.1}$ & $3.30_{-0.02}^{+0.05}$ & $317.83_{-0.33}^{+0.80}$ & 7 & 4.68  \\
MN\,48$^{\dagger}$& LBV &PIONIER & $0.49_{-0.02}^{+0.02}$ & $2.21_{-0.25}^{+0.30}$ & $4.1_{-0.1}^{+0.1}$& $111.91_{-0.24}^{+0.23}$ & $221.98_{-0.13}^{+0.12}$ & 15 & 0.52   \\
MN\,64$^{\dagger}$& cLBV &PIONIER & -- & -- & -- & -- & -- & -- & -- &  no detection\\
WRAY\,17-96& cLBV & PIONIER & -- &  -- & -- & -- & -- & -- & -- &  no detection\\
$\eta$ Car$^{*}$& LBV & PIONIER & -- & -- & -- & $\sim 8$ & $\sim 324$ & -- &-- &  \citet{weigelt07}\\
\hline                                        
\end{tabular}
\tablefoot{
\tablefoottext{a}{From \citet{caballero14} and \citet{maryeva16}: Schulte~12 has a detected companion with a separation $\rho = 65 \pm 1$\,mas, a position angle $PA =  293.0 \pm 0.3^{\circ}$, and with $\Delta\,{\rm mag} = 1.79 \pm 0.02$.}\\
\tablefoottext{$^{\dagger}$} {The names of the stars were shortened for practicality and these objects must be read as ${\rm [GKF2010]}$~MN\,46, ${\rm [GKF2010]}$~MN\,48, and ${\rm [GKF2010]}$~MN\,64.}

}
\end{table*}

To these 11 companions, we have to add 3 objects for which companions were already detected: $\eta$~Car (observed with VLTI/GRAVITY, \citealt{gravity18}), HR~Car \citep{boffin16} and Pistol star \citep{martayan12}. By adding these stars, the astrometric binary fraction rises from $f_{\rm astro} = 0.61$ to $f_{\rm astro} = 0.78$ (14 objects out of 18). Among the 14 companions, seven have an angular separation ($\rho$) between 1 and 10 mas, five have $10 < \rho < 100$ mas, and one has $\rho > 100$ mas (Table\,\ref{table:interf}). Finally, the companion of Schulte~12 must have an angular separation between 65\,mas \citep{caballero14,maryeva16} and about 100\,mas.
Because of the interferometric field-of-view and bandwidth-smearing, PIONIER (resp. MIRC-X, PRISM50 mode) together with the VLTI (resp. CHARA) baselines is hardly sensitive to any binaries with $\rho > 150$ mas (resp. $\rho > 100$ mas).

\begin{figure}[t!]\centering
   \includegraphics[width=9cm,trim=20 0 40 20,clip]{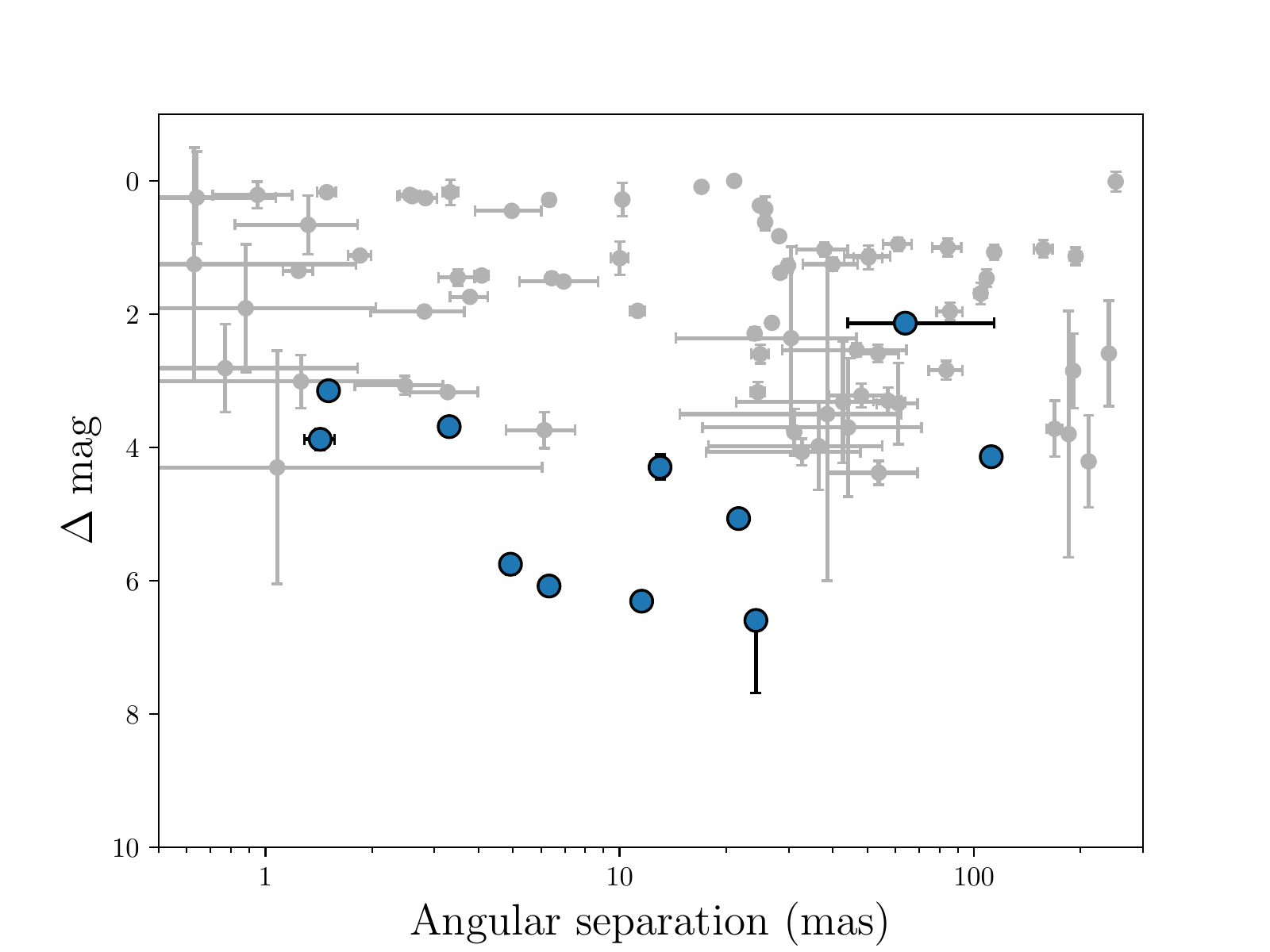}
    \caption{\label{Fig:Deltamag} Plot of the magnitude difference ($\Delta$\,mag) vs. angular separations ($\rho$) for the detected companions. Grey dots correspond to the detected pairs by PIONIER and SAM in the O-type star population \citep{sana14}. }
\end{figure}

For the targets for which there is an overlap in the spectroscopic and interferometric samples, it is unclear whether the detected companions are the same or not. Only dedicated monitorings will help us to understand the real configurations of those systems. No pair with similar or relatively close brightness ($\Delta {\rm mag} < 2$, Fig.\,\ref{Fig:Deltamag}) has been detected in our sample. All the detected companions are much fainter than the companions found around O-type stars \citep{sana14}. In addition, over half of the detected companions have a $\Delta {\rm mag} > 4$ in the $H$ band, i.e., lays beyond the detection limit of the O-star survey. Despite our improved sensitivity, we are only able to detect a maximum of one companion around LBV-like objects while for the O-type star population the average number of companions detected by interferometry is $f_c = 2.1 \pm 0.2$ \citep{sana14}.

The relative flux fractions of the secondaries are converted into flux ratios. From the luminosities of the LBVs \citep[][see also Table\,\ref{table:param}]{smith19,bailerjones21}, these flux ratios provide us with information about the nature of the companions. The luminosities that we can derive for the companions range between $2.8 < \log(L/L_{\odot}) < 5.4$ but they obviously also depend whether we consider the distances given by \citet{smith19} or by \citet{bailerjones21}. Even though we do not know the effective temperatures of the companions, such luminosities suggest that companions are OB stars on the main sequence. We cannot reject that these companions could be RSG, although this is less likely given evolutionary considerations. The fact that we do not observe companions with luminosities higher than $\log(L/L_{\odot}) \sim 5.4$ highlights that these companions cannot be classical WR stars formed through single-star channel nor early O-type or WNL stars, i.e., WR stars on the main-sequence (see Sect.\,\ref{subsec:binaryevol}).


\section{Discussion}\label{Sec:discussion}

\subsection{Physical parameters of the LBV-like objects}
\label{subsec:parameters}
\begin{figure}[t!]\centering
   \includegraphics[width=9cm,trim=20 0 40 20,clip]{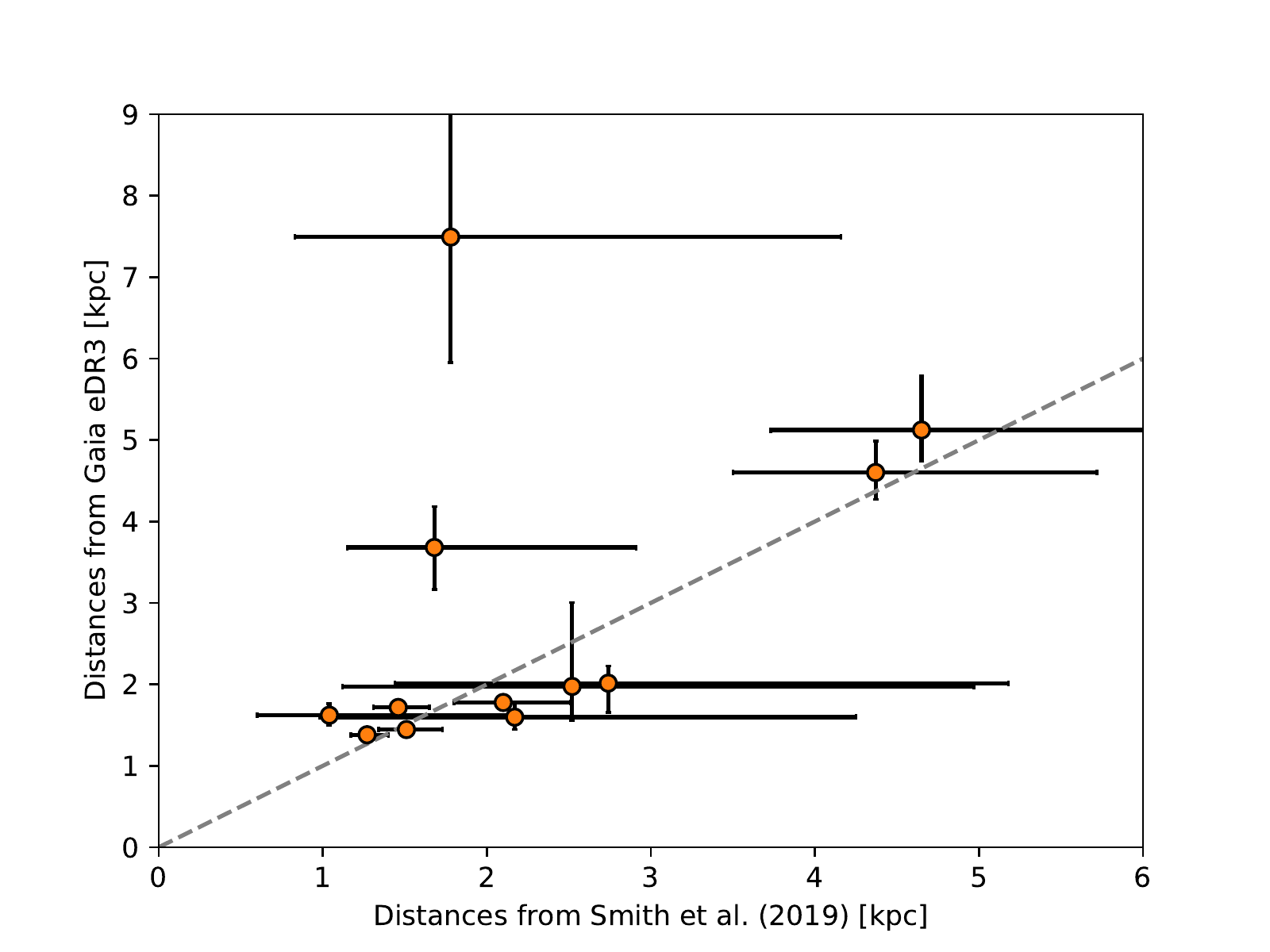}
    \caption{\label{Fig:distances} Comparison between distances from \citet{smith19} and from Gaia eDR3 \citep{bailerjones21}.}
\end{figure}

Using interferometry to characterise the multiplicity of LBV-like objects allows us to derive the primary diameters, and by knowing the distances of the stars, to perform a direct measure of their radii. From our analysis, we have detected companions for 11 objects in our sample. We list these objects in Table\,\ref{table:param}. We also include HR~Car as this star was intensively monitored by \citet{boffin16} and that the primary diameters have been published. 
We retrieve from \citet[][and references therein]{smith19} the luminosities and the effective temperatures of the LBV-like objects from their locations in the HRD. From the luminosities and effective temperatures, we derived their radii (reported as $R_{\rm evol}$, from $L \sim R_{\rm evol}^2 T_{\rm eff}^4$). The radii measured from interferometry are scaled relative to the distances of the stars (reported as $R_{\rm interfero}$ in Table\,\ref{table:param}). Table\,\ref{table:param} summarises the measured radii, luminosities, effective temperatures, distances of the stars reported by \citet{smith19} and \citet{bailerjones21}. We also indicate the size and the shape of their circumstellar nebulae, when they have been detected. Certain distances provided by \citet{smith19} are not in agreement with the distances given by the Gaia early-Data Release 3 \citep[eDR3,][]{bailerjones21} as, e.g., for \west\ and [GKF2010]\,MN46. A comparison is displayed in Fig.\,\ref{Fig:distances}. We stress however that independent studies have shown that, for \west, its real distance is more between 2.6 and 4.1~kpc \citep{Aghakhanloo20,beasor21}. However, considering different distances does affect the measurements of the stellar radii but would not change the conclusions of our study. 

In general, the agreement between the interferometric and evolutionary radii is excellent within the error bars, as it is the case notably for Schulte\,12, [GKF2010]\,MN48, \west, and HR\,Car. For three confirmed LBVs: P~Cyg, HD\,168607, and AG~Car, the radii derived from interferometry are slightly larger than those inferred from the effective temperatures and luminosities. 
One possible explanation is that LBVs can produce optically thick winds around their photosphere, which makes the stars appear bigger than they actually are. Another explanation is that they can present variations of their stellar parameters due to their strong variability or S-Dor cycles that would modify the properties of the stars between the time of our observations and the results listed in \citet{smith19}. AG~Car is an object that is well observed photometrically to probe its variability linked to its S-Dor cycle. When our interferometric observation was collected, AG\,Car was close to its visual maximum phase, i.e., when the star appears cool and eruptive (marked with the red line in Fig.\,\ref{Fig:AGCar_lc}). This phase is characterised by a low effective temperature ($\sim 9$~kK) and a large radius. The radius that we measure from interferometry supports the idea of this eruptive phase that AG~Car went through.



The biggest discrepancy between the radii measured from interferometry and the radii computed from the position in the HRD is observed for HD~326823 (see Table\,\ref{table:param}). HD~326823 was found to be a close binary system with an orbital period of 6.1 days \citep{richardson11}. The companion detected in spectroscopy cannot be the same as that detected from interferometry, meaning that the system consists in a hierarchical triple system. According to \citet{richardson11}, the inner binary system in HD~326823 is surrounded by a circumbinary disk (detected in spectroscopy through double-peaked emission profiles visible, notably, in the \ion{Fe}{ii} and \ion{Ni}{ii} lines). Interferometry does not allow us to resolve the inner system but allows us to model it with its circumbinary material by assuming a disk with a radius of about $120-130\,R_{\odot}$ at the distance of the object. \citet{richardson11} already estimated the inner radius of the circumbinary region to be equal to $2.83 a$ (where $a$ is the semi-major axis of the inner system, i.e., between $30$ and $50\,R_{\odot}$, depending on the inclination of the inner system and the mass of the close-by companion).

We should stress that some objects in our sample are surrounded by large nebulae of ejected material, with sizes of several parsecs squared at their distances (given in Table\,\ref{table:param}). Given the size of these ejecta (much larger than what we measure), it is unlikely that they affect the measurements of the stellar radius in interferometry. These nebulae are also very bright in the infrared (with peaks close to $24\,\mu$m), but remain faint in the $H$-band. Their contributions in our observations are therefore not significant.

Finally, if the radii measured when we consider the distances of the stars are those of the photosphere of the LBVs, this favours the presence of companions on wide orbits.  We note however that we cannot completely rule out the fact that the primaries might be short-period systems surrounded by circumbinary disks as it is the case for HD~326823 but our spectroscopic campaign would have detected 90\% of them.

\begin{sidewaystable*}
\caption{Surrounding environment and physical parameters of the primary}              
\label{table:param}      
\centering                                      
\begin{tabular}{l c | c c c c | c c c c | c c }          
\hline\hline                        
 & & \multicolumn{4}{c|}{Gaia eDR3 } & \multicolumn{4}{c|}{Smith et al.} & \multicolumn{2}{c}{Nebula} \\
\hline
Star & $T_{\rm eff}$ & Distance & $\log(L/L_{\odot}) $ & $R_{\rm interfero}$   & $R_{\rm evol}$ & Distance & $\log(L/L_{\odot}) $ & $R_{\rm interfero}$   & $R_{\rm evol}$ & Size nebula & Shape nebula \\
     & [kK] &  [kpc]  &  &  [$R_{\odot}$]  &   [$R_{\odot}$] &  [kpc]  &  &  [$R_{\odot}$]  &   [$R_{\odot}$] & [pc]  &    \\ 
\hline                                   
AG\,Car & [$7.5-24.0$] & $5.12_{-0.38}^{+0.67}$ & [$5.8-6.2$] & $595_{-44}^{+78}$ & [$45-732$] & $4.65_{-0.92}^{+1.43}$ & [$5.7-6.1$] & $540_{-107}^{+166}$  & [$41-665$] &  $0.95 \times 0.95$  & bipolar\\  [2pt]
Schulte\,12 & [$12.5-14.5$] & $1.62_{-0.12}^{+0.14}$ & [$6.1-6.3$] & $222_{-17}^{+20}$ & [$175-296$] & $1.04_{-0.44}^{+1.17}$ & [$5.7-5.9$] & $142_{-60}^{+160}$  & [$112-190$]  & -- & --  \\[2pt]
HD\,160529  & [$7.0-11.5$] & $1.78_{-0.09}^{+0.07}$ & [$5.2-5.3$] & $216_{-11}^{+9}$ & [$95-289$] & $2.10_{-0.30}^{+0.41}$ &  [$5.3-5.4$] & $255_{-37}^{+50}$ & [$113-341$] & -- & -- \\[2pt]
HD\,168607 & [$9.5-11.5$] & $1.72_{-0.07}^{+0.09}$ & [$5.1-5.2$] & $172_{-7}^{+9}$ & [$94-154$] & $1.46_{-0.15}^{+0.19}$ & [$5.0-5.1$] & $146_{-15}^{+19}$  & [$80-131$] & -- & --  \\[2pt]
HD\,168625 &[$13.0-15.0$] & $1.45_{-0.05}^{+0.05}$ & [$5.0-5.1$] & $-$ & [$45-67$] & $1.51_{-0.17}^{+0.22}$ & [$5.0-5.1$] & $-$   & [$47-70$] & $0.11 \times 0.15$ & bipolar  \\[2pt]
HD\,326823 & [$20.0-22.0$] & $1.38_{-0.04}^{+0.04}$ & [$4.9-5.1$] & $131_{-6}^{+6}$ & [$19-29$]& $1.27_{-0.10}^{+0.13}$ & [$4.8-5.0$] & $120_{-10}^{+13}$ & [$17-26$] & -- & --  \\[2pt]
${\rm [GKF2010]}$~MN~46 & $-$ & $3.68_{-0.52}^{+0.50}$ & $-$ & $554_{-78}^{+76}$ & $-$ & $1.68_{-0.53}^{+1.23}$ &  $-$ & $253_{-80}^{+185}$ & $-$ & $0.49 \times 0.49$ & spherical \\[2pt]
${\rm [GKF2010]}$~MN~48 & [$13.0-15.0$] & $2.02_{-0.36}^{+0.21}$ & [$5.4-5.5$] & $106_{-19}^{+12}$ & [$77-115$] & $2.74_{-1.30}^{+2.44}$ & [$5.7-5.8$] & $144_{-69}^{+129}$ & [$105-157$] & $1.13 \times 1.13$ & spherical \\[2pt]
P\,Cyg & [$18.5-20.5$] & $1.60_{-0.15}^{+0.19}$ & [$5.7-5.9$] & $131_{-12}^{+16}$ & [$58-90$] & $2.17_{-1.19}^{+2.08}$ & [$6.0-6.2$] & $177_{-97}^{+170}$ & [$79-123$]  & $0.09 \times 0.09$ & spherical    \\[2pt]
W\,243 & [$8.0-18.5$] & $7.49_{-1.54}^{+1.56}$ & [$6.1-6.3$] & $653_{-143}^{+147}$ & [$116-778$] & $1.78_{-0.95}^{+2.38}$ & [$4.9-5.1$] & $155_{-83}^{+208}$ & [$27-185$] & -- & -- \\[2pt]
$\zeta$ Sco & [$17.0-19.0$] & $1.98_{-0.42}^{+1.03}$ & [$6.0-6.2$] & $159_{-34}^{+83}$ & [$91-143$] & $2.52_{-1.40}^{+2.45}$ & [$6.2-6.4$] & $203_{-113}^{+198}$ & [$116-183$] & -- & --  \\[2pt]
HR\,Car & [$9.0-19.0$] & $4.60_{-0.33}^{+0.38}$ & [$5.5-5.6$] & $183_{-28}^{+38}$ & [$55-273$] & $4.37_{-0.87}^{+1.35}$ & [$5.5-5.6$]& $174_{-48}^{+63}$  & [$52-260$]  & $0.17 \times 0.13$ & bipolar    \\[2pt]
\hline                                             
\end{tabular}
\tablefoot{
$T_{\rm eff}$ and $\log(L/L_{\odot}) $ are taken from \citet[][and references therein]{smith19} with $1\sigma$ errors equal to 1000K and 0.1, respectively. $R_{\rm interfero}$ is the radius computed from the primary diameter and the estimated distance of the star while $R_{\rm evol}$ is the radius computed from the effective temperature and luminosity. \\
}
\end{sidewaystable*}


\subsection{LBV versus O star populations}
\label{subsec:LBV_O}

Our analysis of a representative sample of Galactic LBVs in the Milky Way both from spectroscopy and interferometry yields a global binary fraction of 78\%. This fraction agrees with that derived among massive Galactic O-type stars \citep{sana12,sana14}. As emphasised in Sect.\,\ref{subsec:parameters}, one difference lies in the number of short- versus long-period systems. About 60\% of binary systems containing massive OB stars have periods shorter than 10 days and 20\% have orbital periods between 10 and 100 days \citep[][Banyard et al. 2021, submitted, Villasenor et al. 2021, submitted]{sana12,almeida17}. The picture is significantly different among the population of Galactic LBV-like objects since only one (HD~326823) is clearly identified as a short-period system. The fact that there is a lack of short-period systems among LBV-like star population is not surprising since the LBV-like objects have relatively large radii that prevent any close companions to exist (Sect.\,\ref{subsec:parameters}).

Since the secondary components are on long-period orbits, we compare the companions around LBV-like stars with the companions detected through interferometry. We focus on the results provided by the Southern MAssive Star at High angular resolution (SMASH+, \citealt{sana14}) survey. This survey combined long-baseline interferometry with VLTI/PIONIER, aperture masking with VLT/NACO-SAM, and adaptive optics with VLT/NACO-FOV to search for companions with separations in the range of 1 to 8000 mas. We only compare the two populations up to a separation of 120 mas, which is the limit of detection in interferometry. As mentioned in Sect.\,\ref{subsec:interfero}, the luminosities of the companions orbiting around LBV-like stars range between $\log(L/L_{\odot}) = 2.8$ and $5.4$, meaning that they are classified between late B and mid O if they are on the main sequence but can also be more evolved and be classified as yellow or red supergiants, if they left the main sequence. From SMASH+, the spectral types of the companions range between $\log(L/L_{\odot}) = 3.4$ and $5.7$, i.e., between mid B and early O. Given that the most massive stars in these systems are on the main sequence, it is more likely that their companions are also located on the main sequence. Therefore, if we assume that the companions around LBVs are still on the main sequence, they are not significantly less massive than the companions around O-type stars. 

\begin{figure}[t!]\centering
   \includegraphics[width=9cm,trim=10 0 40 20,clip]{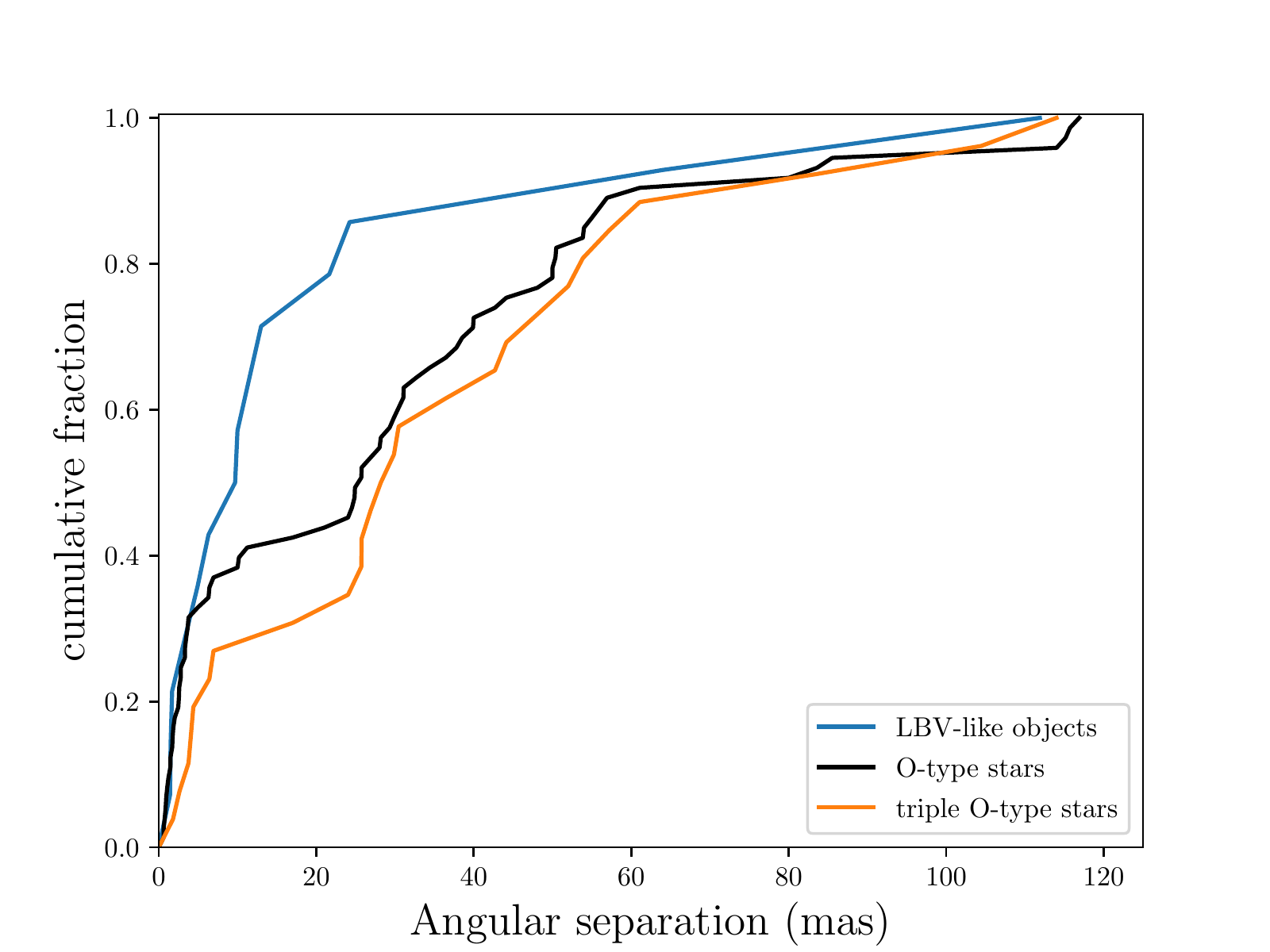}
   \includegraphics[width=9cm,trim=10 0 40 20,clip]{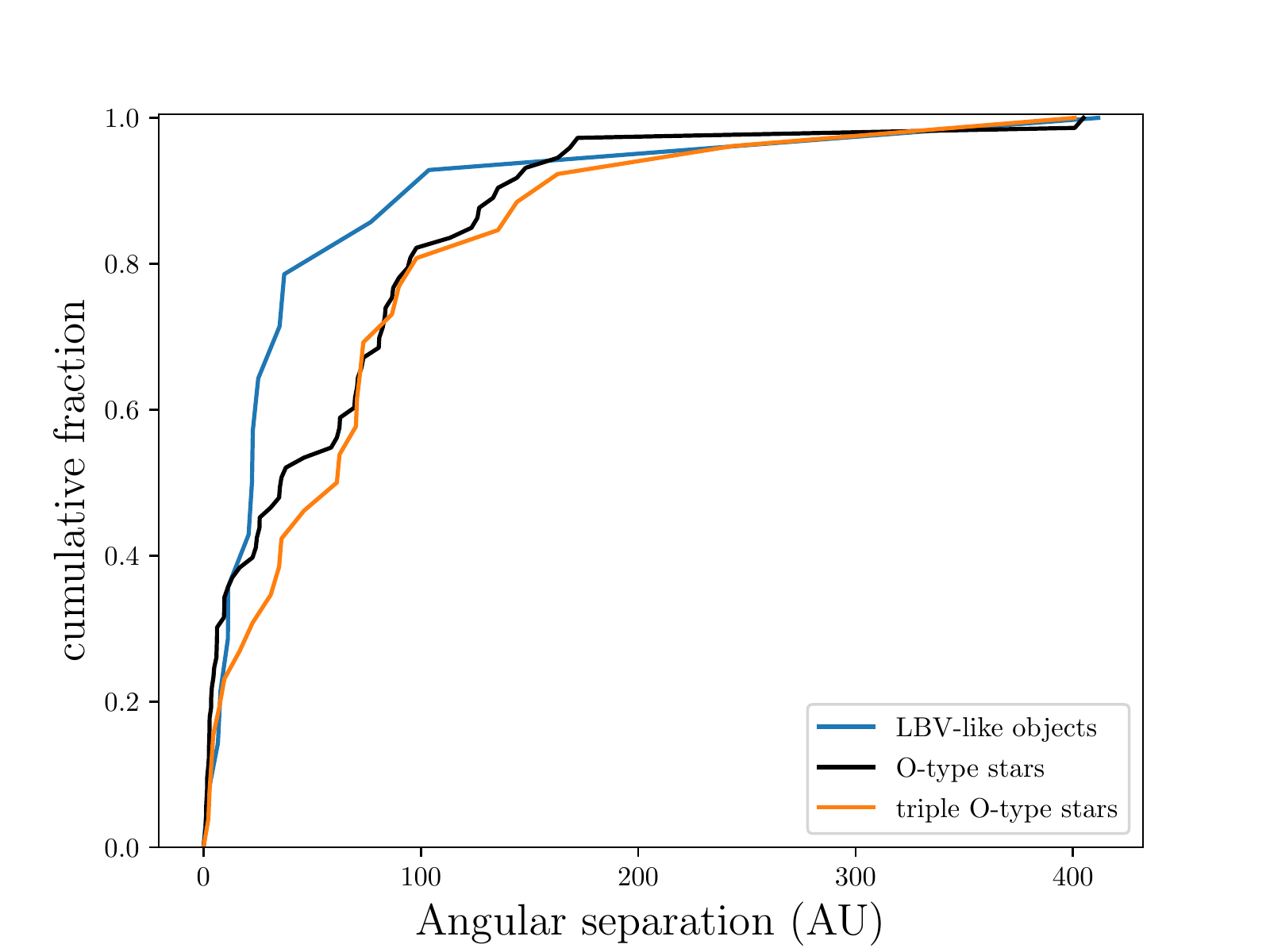}
    \caption{\label{Fig:cumulative} Top: Cumulative distribution of the companion separations among the LBV-like star population (in blue) and among the O-type star population detected through the SMASH$+$ survey \citep{sana14}. Bottom: Same as top panel but considering the distances of the stars from the eDR3 Gaia catalogue.}
\end{figure}

The distribution of angular separations of the companions in the LBV-like star population are compared to those found around O-type stars (top panel of Fig.\,\ref{Fig:cumulative}) by performing a Kuiper test on the two samples. This test indicates that there are only 4\% chances that the two samples come from the same parent distribution ($p-{\rm value} = 0.04$). About 80\% of the LBV-like stars have a companion closer to 20~mas while this value drops to approximately 60\% in the O-star sample. If we only consider the hierarchical triple systems in the O-star sample, i.e., objects composed of an inner short-period spectroscopic binary and an outer star orbiting around it, the angular separation distribution also looks different, we can also reject the null hypothesis that the two distributions are drawn from the same parent distribution ($p-{\rm value} = 0.03$).

Once we take the distances of these systems into account (bottom panel of Fig.\,\ref{Fig:cumulative}), the separation distributions still look different. We perform a Kuiper test to compare the separation distribution of the LBV-like stars and that of the whole sample of O-type stars. We reject the null hypothesis that the two samples come from the same parent distribution ($p-{\rm value} = 0.05$). When we focus only on the hierarchical triple systems, the test also shows that there are 24\% chances that the two distributions are compatible ($p-{\rm value} = 0.24$). We emphasise that the comparison between the triple systems and the whole sample of O-type stars shows that the two distributions are also compatible ($p-{\rm value} = 0.99$). Yet, there are about 80\% of the LBV-like stars that have a companion closer than 50~AU, while this number drops to 53\% among the O-star population (against 47\% if we only consider the triple systems). No matter whether LBV-like stars are formed through single- or binary-evolutionary channel, one explanation to this difference between the distributions could come from the fact that the systems hosting LBV-like stars are more evolved than those hosting O-type stars. However, only a good knowledge of the orbital parameter distributions (orbital period, eccentricity and mass ratio) would give us insight about the role of the LBVs in the evolution of massive stars. 

\subsection{Binary evolution}
\label{subsec:binaryevol}

The traditional view of massive star evolution is that massive O-type stars evolve to classical WR stars (hydrogen-free) by stripping their outer layers via their powerful stellar winds. With the detection of inhomogeneities (or clumps) in their wind, these mass-loss rates were considerably revised downwards among the massive star population \citep{everberg98,vink01,bouret03,bjorklund20} and, when integrated over the stellar lifetime of these stars, they are not sufficient to enable direct evolution between the massive stars on the main sequence and the WR phase. The LBV evolutionary phase comes to justify the transition between these two evolutionary stages. It is thought to be extremely brief  ($10^4 - 10^5$ yrs) in comparison with the lifetime of a massive star ($\sim 10^7$ yrs) and that is during this time span that their nebulae are expected to form through giant eruptions \citep{mahy16}.

 If we consider single-star evolution to explain the existence of LBVs, LBVs must only form in long-period systems so that there is no interaction with their companion during their evolution. Indeed, in these systems, the two components are expected to evolve like single stars most of their life. There are 68 objects that were reported as confirmed or candidate LBVs in the catalogue of \citet{naze12}. Considering the lifetime of the LBV phase, and the number of O-type stars in the Milky Way \citep[][]{maiz11}, about 1\% of the O stars will enter into a phase of LBV. This is less than the number of O-type stars that are expected to evolve like single stars either isolated or as members of binary systems with wide orbits \citep{sana12,demink14}. Therefore, we cannot exclude that LBVs do indeed form through single-star channel. 
  
\begin{figure}[t!]\centering
   \includegraphics[width=9cm,trim=5 0 40 20,clip]{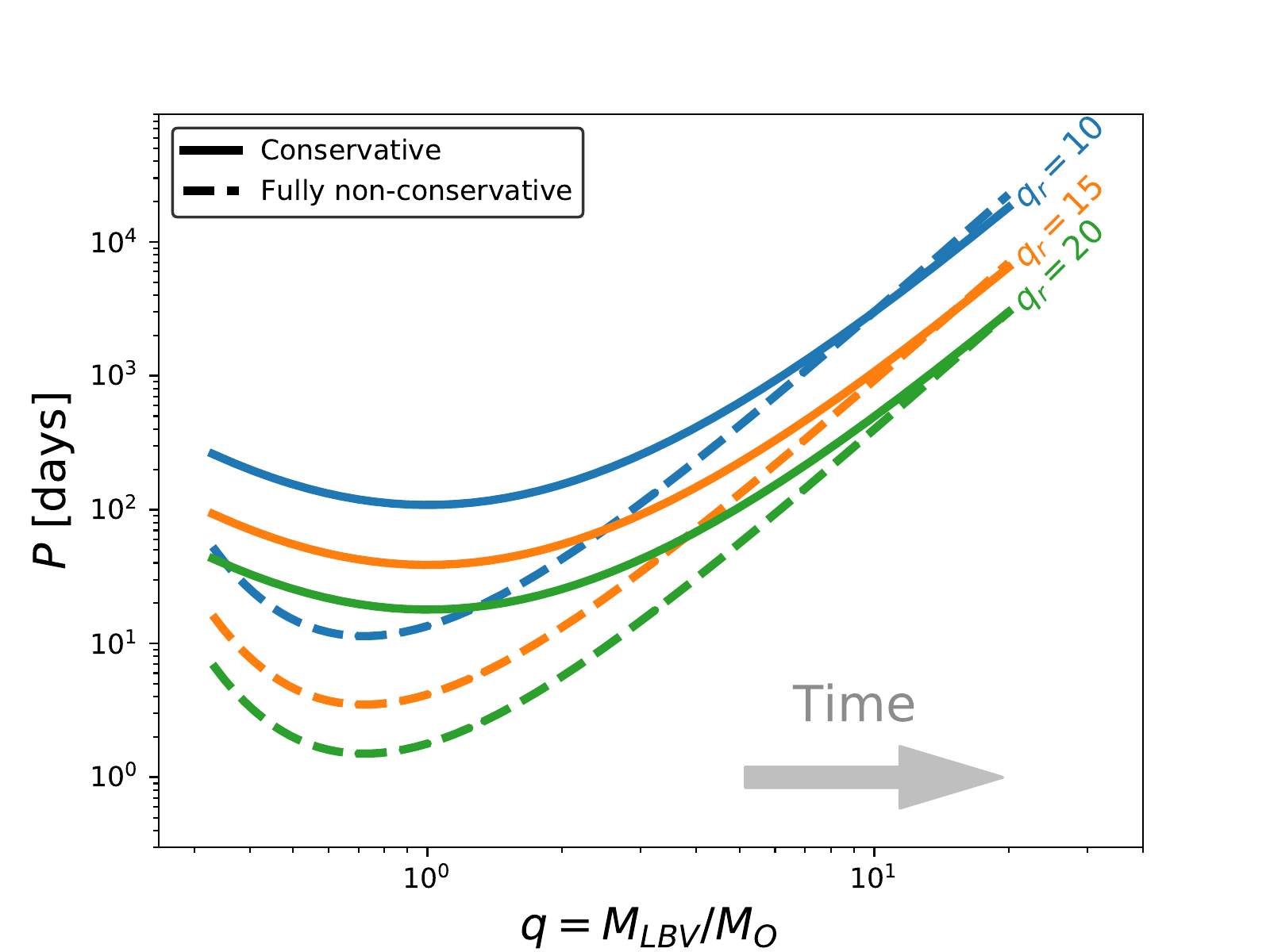}
    \caption{\label{Fig:conservative_diagram} Past orbital evolution of systems assuming present-day mass ratios of $q = 10, 15$ and $20$. During mass transfer phases time increases towards the right in this Figure. Each curve has been computed assuming a current orbital period of 3000 days. The left-most point corresponds to an arbitrary choice of $q = 1/3$.}
\end{figure}
 
 While 30\% of O-stars on the main-sequence are expected to evolve like single stars, most of the stars born as O-type stars will interact during their lifetime. To know whether the abundance of short-period systems among the O-star population ($\sim 60$\%) might be reconcilable with the fact that the majority of LBV-like stars are on wide-orbit systems, we focus on the same approach as presented in \citet{bodensteiner20}. In this approach, the current orbital properties of LBVs ($P_{\rm orb} > 1000 \, {\rm days}$, and mass ratio $q = (M_{\rm LBV}/M_{\rm comp}) < 20$) can be used to assess the initial conditions of the progenitor systems. Assuming circular orbits, a constant mass transfer efficiency, and ignoring other mechanisms for angular momentum loss from the system, the orbital period can be computed analytically as a function of the mass ratio \citep[we refer to ][for the equations]{soberman97,bodensteiner20}. Figure\,\ref{Fig:conservative_diagram} shows the result of evaluating the initial orbital periods and mass ratios for different values of the currently observed mass ratio within our range of uncertainty. We consider two different cases, when the mass transfer is 1) conservative and 2) fully non-conservative.
 In the case of conservative mass transfer and assuming an initial mass ratio of $q = 1/3$, it is unlikely that close binary systems (with initial orbital periods less than 10 days) can evolve to form binary systems with periods of the order of thousand days. When we consider a fully non-conservative mass transfer however such initially tight systems can evolve to form long-period systems ($P_{\rm orb} \gtrsim 10^3$~days) but require to have a current mass ratio higher than 10. Therefore we cannot rule out that detached short-period systems on the main-sequence will evolve to form long-period systems. In this mass-transfer scenario, the initial primary star would undergo a Case A (if still on the main sequence) or an early Case B (post main sequence) mass transfer. This initial primary star, or mass donor, would lose mass and angular momentum and would strip its hydrogen envelope, leading to the formation of a WN star. The initial secondary would accrete this material, inverting the mass ratio. This component, or mass gainer, would gain mass and angular momentum, resulting to a rapid rotator and rejuvenated star. While the mass donor would look like a WR star, the mass gainer could have all the characteristics of LBV stars (e.g., rapid rotation, chemical enrichment, see \citealt{groh09b}). Given the luminosities that we derived for the companions of LBV-like stars (Sect.\,\ref{subsec:interfero}), we cannot exclude that some of these companions, but not all of them, might be stripped WR stars \citep{shenar20}. To this stage of our analysis, we are not yet able to characterise any further the nature of the companions, and whether or not they would have the stellar properties reminiscent of donor stars \citep{mahy20}.
 
 Another scenario to transform short-period systems into wide orbit binaries would involve merging either through a common envelope phase in an initial binary system leading to a single LBV star or in a hierarchical system leading to a binary system with a wide orbit. While we have shown that the separations between main-sequence O-type stars (in binary or triple systems) or LBV-like objects with their companion are not compatible, only the orbital parameter distributions derived for LBV-like stars would allow us to quantitatively test whether these objects might be formed from the binary channel. It is therefore of paramount importance to perform more intensive interferometric monitorings of these stars to derive their orbital parameters.  
 
 Mass transfer and merger were already evoked by \citet{smith15}, and later supported by \citet{aghakhanloo17}, to justify that most of the LBVs are found surprisingly isolated when compared to the O- and WR-star populations. While the results from these analyses were strongly debated, if LBVs are kicked mass gainers, they must appear as runaways (at least for some of them). While massive O-type runaways have been detected in short period binary systems, none has been detected as having companions on wide orbits \citep{sana14}. It would be surprising that, after being ejected from a binary or multiple systems, so many LBVs would still be found in long-period binary systems. This clear dichotomy suggests that the kicked scenario is unlikely.

\begin{figure}[t!]\centering
   \includegraphics[width=9cm,trim=10 0 40 20,clip]{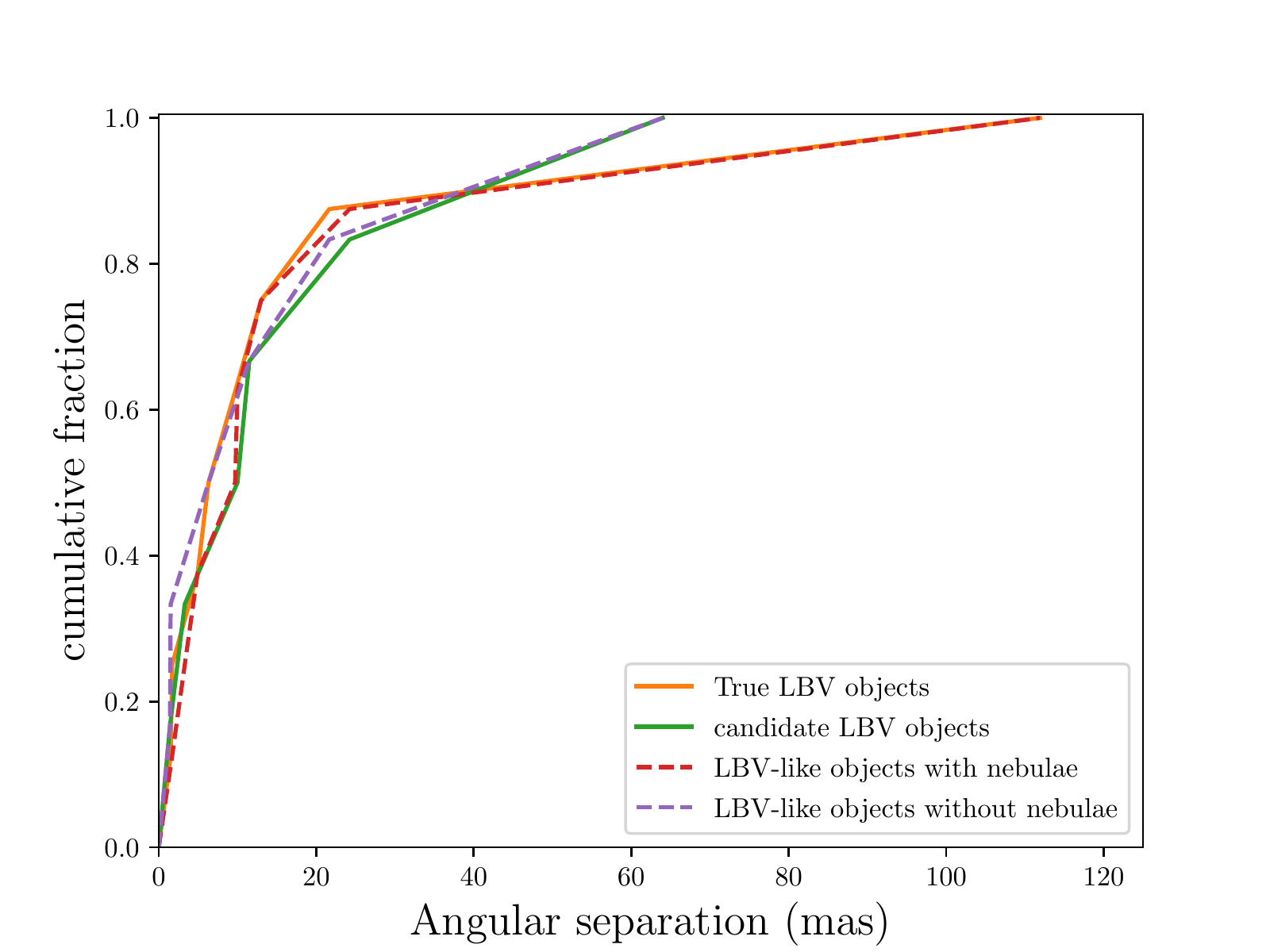}
    \caption{\label{Fig:cumulative_LBV} Cumulative distribution of the companion separations among the LBV-like star population (in blue) and among the O-type star population detected through the SMASH$+$ survey \citep{sana14}.}
\end{figure}

Finally, some (confirmed and candidate) LBVs are surrounded by a circumstellar nebula. The origin of these nebulae is still unknown and might come at this stage from giant eruption, merger and to a lesser extent non-conservative mass transfer. One may wonder whether a link can exist between multiplicity and presence of a nebula. In Table\,\ref{table:param}, we list the morphology and the size of the nebulae detected around LBV-like stars. We display in Fig.\,\ref{Fig:cumulative_LBV} the cumulative distributions of the companion angular separations around LBV-like stars with and without a circumstellar nebula. We cannot reject the null hypothesis that the two distributions are drawn from the same parent distributions ($p-{\rm value} = 0.78$). This may also point out that the kinematics of the outflows in these nebulae deserve to be better studied to verify whether inclination effects may misinterpret the nebulae as spherical rather than bipolar. At this stage, the question about the presence of nebulae around LBVs and possible mass-transfer episodes remains open. We also did not detect any evidence between physical or projected separations of the companions and the size of the nebulae. We also compute a Kuiper test between the populations of confirmed LBVs and candidate LBVs to detect any differences between these two populations and could not find any ($p-{\rm value} = 0.85$). This result however must be taken with caution given the small-number statistics.

\section{Conclusions}\label{Sec:conclusion}
In this study, we performed a dedicated search for binary companions among a Galactic LBV-like star population. We combined high-resolution, high-quality spectroscopic data to interferometric observations to probe separations in the range between 0.1 mas to 100 mas. Our spectroscopic and interferometric samples consist of 18 objects each, and 11 objects appear in both samples.

After having corrected for the spectroscopic observational biases, we found a spectroscopic binary fraction of $62_{-24}^{+38}$\% over the period range between 1 and 1000 days. From our interferometric data, we also find a high binary fraction of 78\%, i.e., for projected separations up to 100-150 mas, depending on the instruments that we use. 90\% of confirmed LBVs in our sample do have a companion, while this fraction goes down to about 45\% for the candidate LBVs. Even though the observations do not allow us to probe the orbital parameter distributions, we measure through interferometry the physical radii of LBV-like stars. These radii suggest that our objects in binary systems must have companions on wide orbit. This significantly contrasts with the O and early B-type stars that are mainly found in systems with orbital periods between 1 and 10 days. We show that we can reproduce the currently observed parameters ($P_{\rm orb} \gtrsim 10^3$ days) with a tight progenitor system that has undergone a full non-conservative Case A mass transfer. Another scenario can also explain the formation of wide-orbit systems: merger in binary or hierarchical triple systems. This challenges the traditional view that O-type stars evolve to WR stars through a transitory LBV phase. 

The interferometric data also allow us to characterise the luminosities of the companions. We estimated that the companions have luminosities between $\log(L/L_{\odot}) = 2.8$ and 5.4, meaning that they are classified from late B to mid O if they are on the main sequence. These luminosities show that early O or WNL star can be excluded as companions of LBV-like stars. Given that we do not have any indications about the effective temperatures of the companions, we cannot rule out that the companions might be RSGs. Additionally, even if it is unlikely, we can also not rule out that they might be classical WR stars. However, the luminosity range that we derived for the companions infers that, if they are classified as WR stars, there are strong indications that they must be formed through binary interactions. 

Our study put for the first time a strong constraint on the binary fraction among Galactic confirmed and candidate LBVs and on the exact radii of these stars. If one want to understand the role of LBVs in the massive-star evolution, additional observations still need to be obtained to characterise the orbital parameters of these newly-discovered systems. These distributions are crucial to quantitatively test the evolution of massive stars.



\begin{acknowledgements}
L.M.\ thanks the European Space Agency (ESA) and the Belgian Federal Science Policy Office (BELSPO) for their support in the framework of the PRODEX Programme. H.S.\ acknowledges support from the FWO\_Odysseus program under project G0F8H6N. The research leading to these results has received funding from the European Research Council (ERC) under the European Union's Horizon 2020 research and innovation programme (grant agreement numbers 772225: MULTIPLES).
J.K. acknowledges support from the research council of the KU Leuven under grant number C14/17/082 and from FWO under the senior postdoctoral fellowship (1281121N). A.L.\ acknowledges funding received in part from the European Union Framework Programme for Research and Innovation Horizon 2020 (20142020) under the Marie Sklodowska-Curie grant Agreement No. 823734 (Physics of Extreme Massive Stars). RM acknowledges funding from the South African Claude Leon Foundation. We are also thankful to Tomer Shenar, Pablo Marchant, and Karan Dsilva for very interesting discussions.
The Mercator telescope, operated by the Flemish Community on the island of La Palma at the Spain Observatory del Roche de los Muchachos of the Instituto de Astrofisica de Canarias. Mercator and Hermes are supported by the Funds for Scientific Research of Flanders (FWO), the Research Council of KU Leuven, the Fonds National de Recherche Scientifique (FNRS), the Royal Observatory of Belgium, the Observatoire de Gen{\`e}ve, and the Th{\"u}ringer Landessternwarte Tautenburg. We thanks all the observers who collected the data with Hermes.
The TIGRE facility is funded and operated by the universities of Hamburg, Guanajuato, and Li{\`e}ge.
This paper is based in part on spectroscopic observations made with the Southern African Large Telescope (SALT) under programme 2019-1-SCI-001 (PI: Miszalski). We are grateful to our SALT colleagues for maintaining the telescope facilities and conducting the observations.
This research has made use of the SIMBAD database, operated at CDS, Strasbourg, France and of NASA's Astrophysics Data System Bibliographic Services. We are grateful to the staff of the ESO Paranal Observatory for their technical support.
\end{acknowledgements}

\bibliographystyle{aa}
\bibliography{LBV_bbl}

\begin{appendix}

\section{Radial velocity measurements}

\begin{sidewaystable*}
\caption{Journal of the spectroscopic observations of the LBV-like stars in our sample. The first column gives the heliocentric Julian date. The following columns provide, for each spectral line/region, the measured RVs (in \kms) and the average values are given in the last column. The asterisk marks the epoch used as template for the cross-correlation. The full table is available electronically at the CDS.}              
\label{table:radvel}      
\centering                                      
\begin{tabular}{l l | c c c c c | c  }          
\hline\hline                        
 Star  & HJD &  \multicolumn{5}{c|}{Individual lines} & Mean RV \\
\hline
AS~314 & & \ion{Si}{ii}~4128-30 & \ion{He}{i}~4471~/~\ion{Mg}{ii}~4481 & \ion{Si}{ii}~6347 & \ion{Si}{ii}~6370 &  &   \\
\hline                                   
& 7599.7293 & $-61.7 \pm   0.9$ & $-62.6 \pm   0.4$ & $-62.4 \pm   1.1$ &$-62.1 \pm   0.9$ & &$-62.2 \pm   0.9$  \\
& 7699.5137 & $-66.3 \pm   1.0$ & $-63.2 \pm   0.6$ & $-57.7 \pm   0.8$ & $-61.6 \pm   1.1$ &&$-62.2 \pm   0.9$ \\
& 9045.5380 & $-49.4 \pm   1.0$ & $-55.8 \pm   0.8$ & $-54.8 \pm   0.7$ & $-54.4 \pm   0.7$ && $-53.6 \pm   0.8$  \\
& 9051.5614 & $-61.5 \pm   1.7$ & $-56.3 \pm   1.4$ & $-58.4 \pm   0.9$ & $-57.8 \pm   1.1$ & &$-58.5 \pm   1.3$  \\
& 9077.4694 & $-44.5 \pm   0.9$ & $-46.9 \pm   0.9$ & $-44.8 \pm   0.8$ & $-44.9 \pm   0.7$ & &$-45.3 \pm   0.8$  \\
& 9111.3504 & $-52.4 \pm   2.1$ & $-47.7 \pm   1.0$ & $-50.5 \pm   1.0$ & $-47.6 \pm   1.1$ &&$-49.6 \pm   1.3$  \\
& 9114.3750 & $-43.8 \pm   0.9$ & $-45.2 \pm   0.9$ & $-46.4 \pm   0.8$ & $-44.8 \pm   0.7$ &&$-45.0 \pm   0.8$  \\
& 9119.3566$^{*}$ & $-38.1 \pm   1.0$ & $-41.8 \pm   0.9$ & $-38.9 \pm   1.0$ & $-39.3 \pm   1.1$ &&$-39.5 \pm   1.0$  \\
& 9122.3463 & $-44.5 \pm   0.8$ & $-42.3 \pm   0.8$ & $-36.7 \pm   0.9$ & $-37.5 \pm   1.0$ && $-40.2 \pm   0.9$ \\
\hline    
HD~160529 & & \ion{Si}{ii}~4128-30 & \ion{Mg}{ii}~4481 & & & &  \\ 
\hline          
&2006.4228 &  $-22.1 \pm   0.3$ & $-22.8 \pm   0.4$ & & & & $-22.5 \pm   0.3$ \\&2351.3948 &  $-21.0 \pm   0.4$ & $-22.5 \pm   0.4$ & & & & $-21.7 \pm 0.4$ \\
&3130.4256 &  $-25.4 \pm   0.3$ & $-26.7 \pm   0.5$ & & & & $-26.1 \pm 0.4$ \\
&3481.3844$^{*}$ &  $-19.7 \pm   0.3$ & $-18.7 \pm   0.5$ & & & & $-19.2 \pm 0.4 $\\
&6421.3867 &  $-20.2 \pm   0.2$ & $-24.9 \pm   1.6$ & & & & $-22.6 \pm 0.8 $ \\
&7313.0194 &  $-21.1 \pm   0.5$ & $-17.1 \pm   1.4$ & & & &  $-19.1 \pm 0.9 $ \\
\hline

\end{tabular}
\end{sidewaystable*}
\end{appendix}
\end{document}